\documentclass[fleqn,usenatbib]{mnras}

\usepackage{newtxtext,newtxmath}
\usepackage[T1]{fontenc}

\DeclareRobustCommand{\VAN}[3]{#2}
\let\VANthebibliography\thebibliography
\def\thebibliography{\DeclareRobustCommand{\VAN}[3]{##3}\VANthebibliography}

\usepackage{graphicx}% Including figure files
\usepackage{amsmath}	% Advanced maths commands
\usepackage{amssymb}	% Extra maths symbols
\usepackage{caption}
\usepackage{subcaption}
\usepackage{booktabs}
\usepackage{longtable}
\usepackage{csvsimple}
\usepackage{dblfloatfix}
\newcommand{\abs}[1]{\left\lvert#1\right\rvert}

\title[Discovering properties of Ca II Absorbers]{Newly discovered Ca II absorbers in the early universe: statistics, element abundances and dust}

\author[Fang et al.]{
Hannah Fang$^{1,2}$\thanks{E-mail:hannahrita.fang@gmail.com}, Iona Xia$^{1,3}$, Jian Ge$^{4}$\thanks{E-mail:jge@shao.ac.cn}, Kevin Willis$^{1}$ and Yinan Zhao$^{5}$
\\
$^{1}$ Science Talent Training Center, Gainesville, FL 32606, USA \\
$^{2}$ Centennial High School, Ellicott City, MD 21042, USA \\
$^{3}$ Monta Vista High School, Cupertino, CA 95014, USA \\
$^{4}$ Division of Science and Technology for Optical Astronomy, Shanghai Astronomical Observatory, Chinese Academy of Sciences,
Shanghai 200030, China \\
$^{5}$ Department of Astronomy, University of Geneva, Geneva, Switzerland
}

\pubyear{2021}

\begin{document}
\label{firstpage}
\pagerange{\pageref{firstpage}--\pageref{lastpage}}
\maketitle

%abstract revisions are only rewording of sentence structure to cut abstract to below 150 words
\begin{abstract}
We report discoveries of 165 new quasar Ca II absorbers from the Sloan Digital Sky Survey (SDSS) Data Release 7 and 12. Our Ca II rest frame equivalent width distribution supports the weak and strong subpopulations, split at ${W}^{\lambda3934}_{0}=0.7 $\AA. Comparison of both populations' dust depletion shows clear consistency for weak absorber association with halo-type gas in the Milky Way (MW) while strong absorbers have environments consistent with halo and disc-type gas. We probed our high redshift Ca II absorbers for 2175\AA\ dust bumps, discovering 12 2175\AA\ dust absorbers (2DAs). This clearly shows that some Ca II absorbers follow the Large Magellanic Cloud (LMC) extinction law rather than the Small Magellanic Cloud extinction law. About 33\% of our strong Ca II absorbers exhibit the 2175\AA\ dust bump while only 6\% of weak Ca II absorbers show this bump. 2DA detection further supports the theory that strong Ca II absorbers are associated with disk components and are dustier than the weak population. Comparing average Ca II absorber dust depletion patterns to that of Damped Ly$\alpha$ Absorbers (DLAs), Mg II absorbers, and 2DAs shows that Ca II absorbers generally have environments with more dust than DLAs and Mg II absorbers, but less dust than 2DAs. Comparing 2175\AA\ dust bump strengths from different samples and also the MW and LMC, the bump strength appears to grow stronger as the redshift decreases, indicating dust growth and the global chemical enrichment of galaxies in the universe over time.

\end{abstract}


%%%%%%%%%%%%%%%%%%%%%%%%%%%%%%%%%%%%%%%%%%%%%%%%%%

%%%%%%%%%%%%%%%%% BODY OF PAPER %%%%%%%%%%%%%%%%%%

\section{\fontsize{12}{15}\selectfont Introduction}
Spectroscopy is a powerful tool for studying the evolution and properties of distant galaxies. By observing quasar absorption line spectra, we can study the chemical abundances, kinematics, and physical processes of galaxies hosting high redshift quasar absorbers \citep*{2006MNRAS.367..211W}. Through analysis of these observations, we can further study and trace the distribution of gas and dust in galaxies and the intergalactic medium at various redshift ranges \citep*{1965ApJ...142.1677B, 1966ApJ...144..447B}. Since high redshift quasars are among the brightest objects in the early universe, observations of quasar absorption lines provide a wealth of information on the evolution of matter in the universe and details in gaseous structures and physical processes in high redshift galaxies.

 Damped Ly$\alpha$ absorbers (DLAs) and Mg II absorbers are the two popular ones due to larger attainable samples.  DLAs are systems with large H I  column densities of at least $2\times10^{20} \text{ atoms cm}^{-2}$ \citep{1986ApJS...61..249W, Coles2003}. DLAs are important to trace global neutral gas in the early universe. DLAs have been detected in quasar Ca II absorbers as the presence  of Ca II usually indicates a large H I column density.

Historically, Ca II, which features two strong transitions present in the near-UV and optical regime that are observable at low redshifts, has been studied less due to its rarity and prior technological limitations \citep*{2018A&A...609A..85G}. Due to an ionization potential lower than H I (at 11.9 eV), Ca mainly exists as doubly ionized (Ca III) in the interstellar medium (ISM) \citep{2016A&A...591A.137G}. Previous studies by \citet{1996ARA&A..34..279S} found Ca to be highly depleted in the ISM, with over 99 per cent 'hidden' through incorporation into dust grains. The singly ionized Ca II (Ca$^+$) $\lambda \lambda$ 3934, 3969 doublet corresponds to the Fraunhofer H ($\lambda 3969$) \& K ($\lambda 3934$) lines. Ca in the Milky Way has been further studied, serving as a basis of comparison and modeling for disk and halo environments of our galaxy. Ca II comes to light as a new way to find out more about star formation due to its dusty nature (in comparison to the general DLA population and strong Mg II absorbers), host of molecular hydrogen, and its ability to be observed from a low redshift \citep{2006MNRAS.367..211W, 2008MNRAS.390.1670N, 2007MNRAS.379.1409Z, 2015MNRAS.452.3192S}. Ca II absorbers are also the main absorbers with the ability access environments with a redshift that is less than 0.4 (most recent 4.3 billion years), allowing Ca II absorbers to be able to study cool, metal and dust-rich gas around galaxies with lower redshifts.  This can therefore help probe cool gas and dust components in both disk and  halo of galaxies \citep{2014MNRAS.444.1747S}.

Ca II absorbers are often compared to the much more common Mg II absorbers, and Ca II absorbers are often discovered through Mg II absorbers. In the study of Ca II absorbers by \citet{2005MNRAS.361L..30W}  to observe dust in DLAs, they required candidate systems to possess Mg II $\lambda \lambda$ 2796, 2804 absorption line detection. They detected 31 Ca II absorbers and observed that large Ca II 3934 equivalent width ($W_0^{\lambda3934}$) generally has large color excess, $E(B-V)$, than small $W_0^{\lambda3934}$. When they split their sample by $W_0^{\lambda3934}=0.7$ \AA, they measured $E(B-V)=0.099$ and 0.025 for the $W_0^{\lambda3934}\geq0.7$ \AA\ and $W_0^{\lambda3934}<0.7$ \AA\ composite spectra, respectively. \citet{2008MNRAS.390.1670N} also used Mg II absorbers in selecting Ca II absorbers in their study of Ca II absorption in DLAs.

\citet{2006MNRAS.367..211W} found Ca II absorbers in the high $W^{\lambda3934}_0$ regime to be amongst the dustiest absorbers known while those in the lower $W^{\lambda3934}_0$ regimes had smaller levels of depletion. \citet{2007MNRAS.379.1409Z} found 18 of their 19 Ca II quasar absorbers (all in the lower $W^{\lambda3934}_0$ regime) to have <0.3 dex variation in [Cr/Zn], which is consistent with the general DLA population. The less dusty nature of low $W^{\lambda3934}_0$ absorbers also follows the less dusty nature of DLAs. 

Study of Fe II absorption-line profiles of Ca II absorbers suggested environment systems that are not simple disk-like and with general elemental depletion patterns similar to the warm halo phase of the Milky Way. The study by \citet{2013ApJ...770..130Z} suggests that more than 90\% of Ca II in the universe is in the circum- and inter-galactic environments. They also found that more Ca II is present in the halo of star-forming galaxies than quiescent one (at fixed stellar mass), implying that star formation, or a correlated process, is responsible either for injecting Ca II into the halo or maintaining a large amount of calcium in a certain physical condition in the halo. \citet{2013ApJ...770..130Z} also found a correlation between the amount of Ca II in the halo and stellar mass of galaxies. 

In the first large scale study of Ca II absorbers with a catalog of 435 doublets, \citet{2014MNRAS.444.1747S} analysed Ca II absorbers  with redshift coverage of z<1.34 and supported the suggestion that two distinct populations split Ca II absorbers at $W^{\lambda3934}_0 = 0.7$ \AA. The only pre-existing large catalog (>50), is also skewed towards the lower redshift regime with $<z_{\text{abs}}>=0.58$. In a follow-up study by  \citet{2015MNRAS.452.3192S}, element abundances were measured to prove physical differences between the two populations with indications that the weak population has properties consistent with halo-type gas while the strong population has those consistent with halo- and disc-type gas. They also confirmed \citet{2006MNRAS.367..211W}'s results of a greater E(B-V) for the strong population. 

However, \citet{2015MNRAS.452.3192S} could not distinguish whether the LMC or SMC dust-extinction law better fit strong Ca II absorbers based on their data and they concluded consistency with both. A differentiating feature between the LMC and SMC extinction curve, however, is the exhibition of the 2175 \AA\ dust bump. The only Ca II absorber study to consider the 2175 \AA\ dust feature was \citet{2005MNRAS.361L..30W}, and only in the context of ruling out the MW extinction curve. They found the bump feature to be either weak or absent from the spectra of their catalog's Ca II absorbers and thus did not go on to measure bump strength nor probe for dust absorbers. However, their sample was limited to 31 absorbers, only 13 of which were "strong" Ca II absorbers.

We seek to explore the dust content of Ca II absorbers through search for the 2175 \AA\ dust bump. Discovery of significant dust bump features will point towards further Ca II absorber consistency with the LMC dust-extinction law and evaluate the effectiveness of Ca II absorbers as a dust absorber probe. Our study's ability to access a large sample of Ca II absorbers at high redshifts (i.e., z>0.7) will further increase our ability to search for the dust bump, which has only been found at z$\gtrsim0.7$ using ground-based optical spectral data \citep{2004ApJ...609..589W, 2013AJ....145..157J}.

Further, only a limited Ca II absorber sample is currently available because of their rarity in both number and available studies. Thus, a greater number of absorbers, especially those in the higher redshift range, must be discovered to support prior claims and studies regarding their properties and environments.

This paper is organized as follows. In Section 2, we give a description of our methods in identifying and verifying new Ca II absorbers in SDSS DR7 and DR12 data, spectrum stacking to create a composite spectrum for measuring chemical abundances of various species covered by the SDSS quasar spectra, and also searching for 2175 \AA\ dust bumps in Ca II absorbers. In Section 3, we present our results from this study. We conclude this investigation in Section 4, followed up with discussions in Section 5. 

\section{\fontsize{12}{15}\selectfont Methods}
\subsection{Discovering Ca II absorbers}\label{sssec:discovery}
We use quasar spectral data from SDSS-II DR7 \citep{2009ApJS..182..543A} and SDSS-III Baryon Oscillation Spectroscopic Survey's (BOSS) DR12 \citep{2015ApJS..219...12A} for this study. Although prior studies conducted by \citet{2014MNRAS.444.1747S} searched DR7 and DR9 for Ca II absorbers, we re-search DR7 with a different detection method that allows us to discover more Ca II absorbers at high redshifts along with other Ca II absorbers with $z\geq0.36$ that may have been missed by \citet{2014MNRAS.444.1747S}. We search DR12 instead of DR9 as DR12 is a more recent release than DR9 and covers all DR9 data. This allows us to find new Ca II absorbers not included in the DR9 data. 

As Ca II absorbers are relatively weak, Ca II absorbers are not easy to be discovered alone, prompting the use of Mg II absorbers with z < 1.4 as an aide in discovering and verifying Ca II absorbers. Essentially all Ca II absorbers show associated Mg II absorption. \citet{2014MNRAS.444.1747S} found that 251 of the 303 Ca II absorbers in their catalog at $z \geq 0.36$ have associated Mg II absorption and \citet{2011A&A...528A..12R} found their entire catalog of low redshift (z < 0.5) weak Ca II absorbers show associated absorption by Mg II. This practice of requiring a candidate to have Mg II $\lambda \lambda$ 2796, 2804 absorption before being verified as a Ca II $\lambda \lambda$ 3934, 3969 absorber has been used in past Ca II studies \citep{2006MNRAS.367..211W, 2007MNRAS.379.1409Z}.

Mg II absorbers identified in DR7 and DR12 \citep{2013ApJ...770..130Z, 2019MNRAS.487..801Z} were utilized.  These Mg II absorbers were discovered with both Mg II $\lambda\lambda2796, 2803$ lines having an SNR $\geq$ 3. The catalog contained $\sim$35,000 and $\sim$41,000 Mg II absorbers, from SDSS DR7 and DR12 respectively. However, in total, we only had around 43,000 Mg II absorbers with $z_{abs}<1.4$ to search for significant Ca II absorption. These Mg II absorbers served two primary roles: 1) as redshift signposts and 2) to limit the amount of spectra searched. The $z_{abs}$ of each SDSS quasar spectra searched was already calculated, resulting in easier preprocessing of data. This pre-restricted quasar sample eliminated the need for target selection and pre-detected data reduction beyond elimination of absorbers with $z>1.4$. The catalog ensured adequate signal-to-noise ratios (SNR) as the entire quasar sample has SDSS magnitudes i < 20 and that systems were quasars and not misidentified galaxies. 

We created a detection program written in python to measure the SNR of the Ca II $\lambda\lambda3934, 3969$ lines in each spectrum searched and flag those with SNR $\geq3$ and $\geq2.5$, respectively. Our detection program flagged about 1050 spectra for visual inspection to distinctify between real Ca II absorbers, questionable candidates, and false positives. Our detection program first fit a continuum to the spectrum using a median filter. The spectrum was then normalized and a Gaussian model (utilizing \texttt{LMFIT}) was used to fit the features at $\lambda \lambda$ 3934, 3969.  $W_0$ was calculated by $W_0 = A/I$ in which A is the area of the Gaussian, $A = \abs{a\sigma\sqrt{2\pi}}$ (\emph{a} and $\sigma$ are the depth and standard deviation of the Gaussian profile, respectively), and I is the continuum intensity. $W_0$ error, $W_0(err)$ was calculated by multiplying the mean of a spectral error array that corresponds to 2.5 \AA\ around the line center's wavelength by the full width at half maximum (FWHM) of the Gaussian profile.
\begin{figure}
    \centering
    \includegraphics[width =\linewidth]{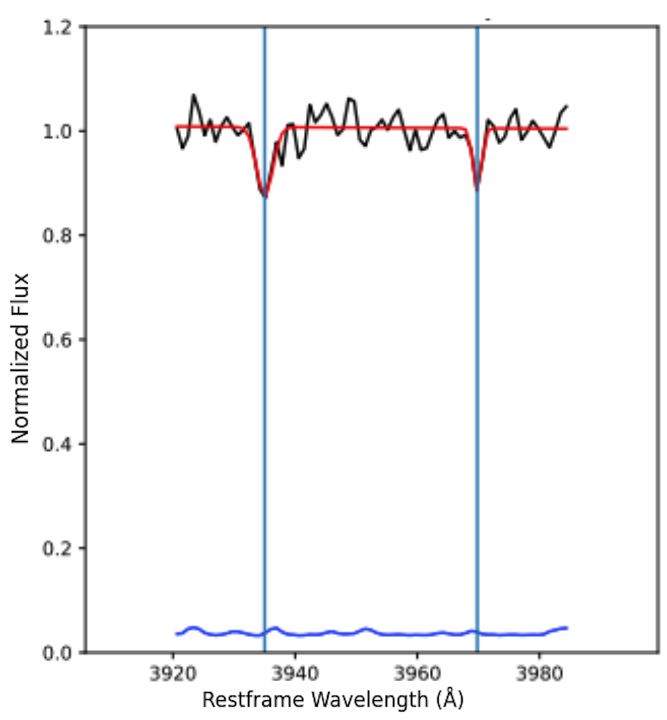}
    \caption{An example of a Ca II candidate flagged by our detection program. The initial Gaussian fits for the $\lambda\lambda3934, 3969$ absorption lines are shown in the red line. The dark blue line near the bottom is the plotted error array.}
    \label{fig:fig1}
\end{figure}

The program searched for Ca II absorption by measuring the line significance levels at the H \& K doublets. To measure line significance, $W_0^{\lambda3934}$ and $W_0^{\lambda3969}$ were measured and their corresponding errors were derived. A quasar spectrum fit by the detection program is shown in Figure \ref{fig:fig1}. All absorbers with SNR = $W_0/W_0(err) \geq$ 3 for the Ca II 3934 line and SNR $\geq 2.5$ for the Ca II 3969 line were flagged as possible Ca II absorber candidates. As the Ca II 3934 line is the more prominent line of the doublet, with an f-value of 0.6346 compared to Ca II 3969's f-value of 0.3145, it requires a greater SNR threshhold for detection.

Our detection program was also successful in recovering Ca II absorbers from \citet{2014MNRAS.444.1747S}'s catalog. 303 of the Ca II absorbers in \citet{2014MNRAS.444.1747S}'s catalog were within our Mg II absorber catalog's redshift range of $\geq0.36$. These 303 absorbers were composed of 146 strong (72\% of all the strong absorbers in \citealt{2014MNRAS.444.1747S}'s catalog) and 157 weak absorbers (67\% of all the weak absorbers in \citealt{2014MNRAS.444.1747S}'s catalog), which means that the redshift range searched is not biased towards either subpopulation. Of those 303 possible Ca II absorbers, 241 were in the Mg II absorber catalog we searched. Of the 62 absorbers not within the searchable Mg II catalog, 53\% were strong Ca II absorbers and 47\% were weak Ca II absorbers, which is not a significant bias for or against either subpopulation. This proves that our method of using Mg II absorbers as a probe is not inherently biased towards finding weak Ca II absorbers.

\begin{figure*}
    \includegraphics[width =.45\linewidth]{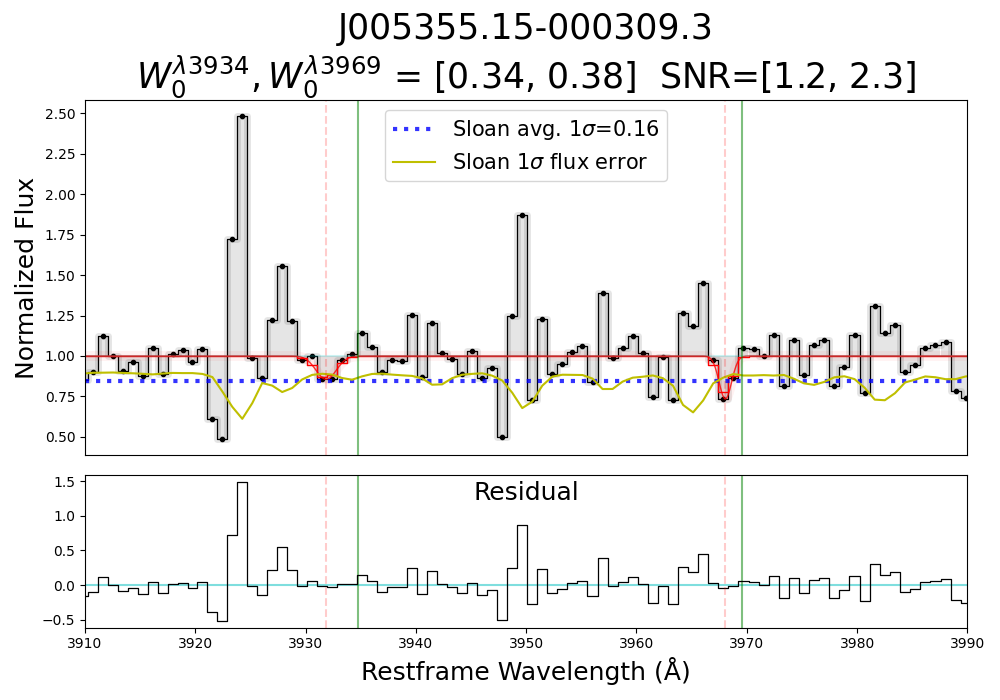}
    \includegraphics[width =.45\linewidth]{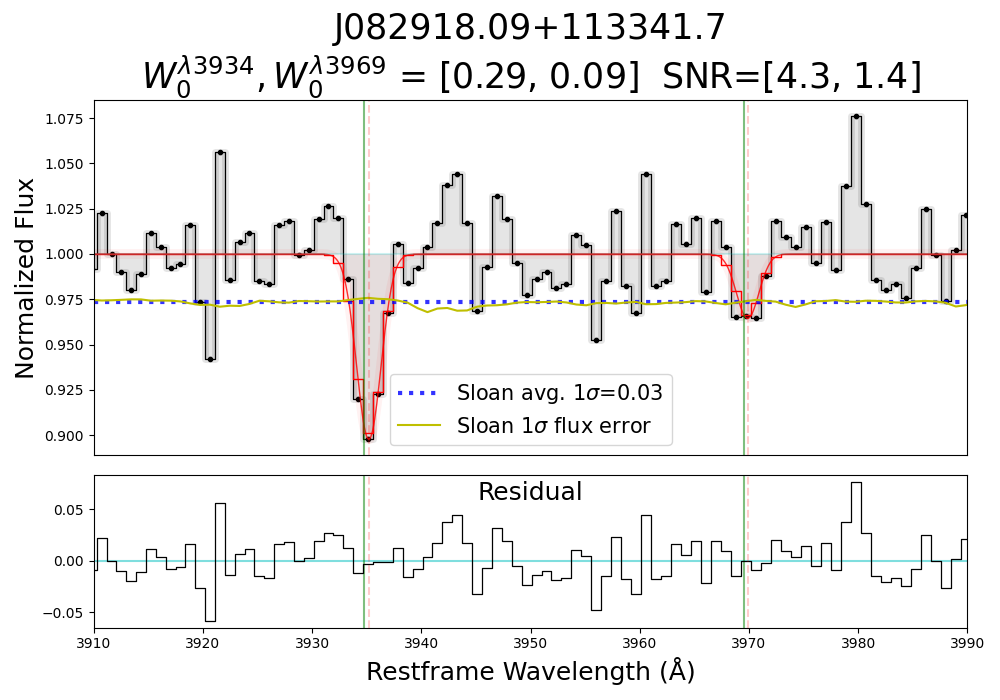}
    \includegraphics[width =.45\linewidth]{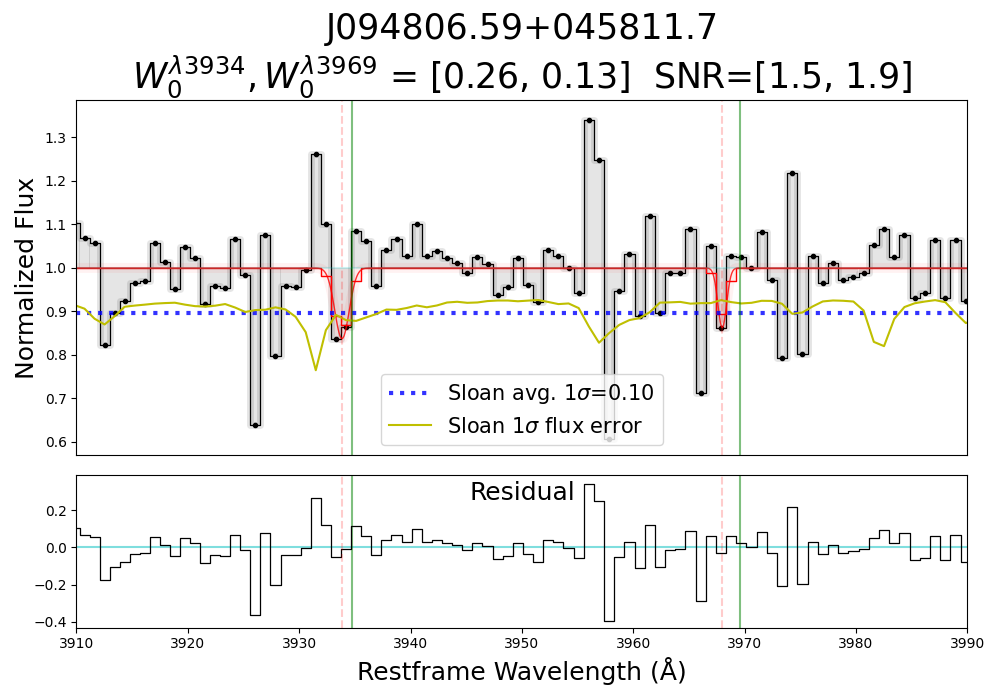}
    \includegraphics[width =.45\linewidth]{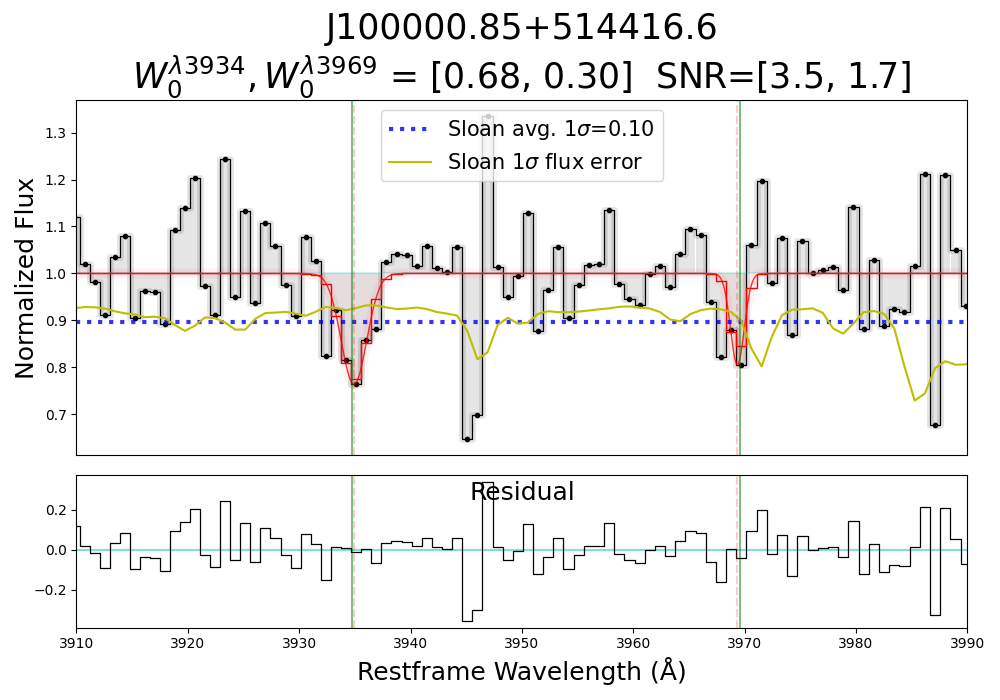}
    \includegraphics[width =.45\linewidth]{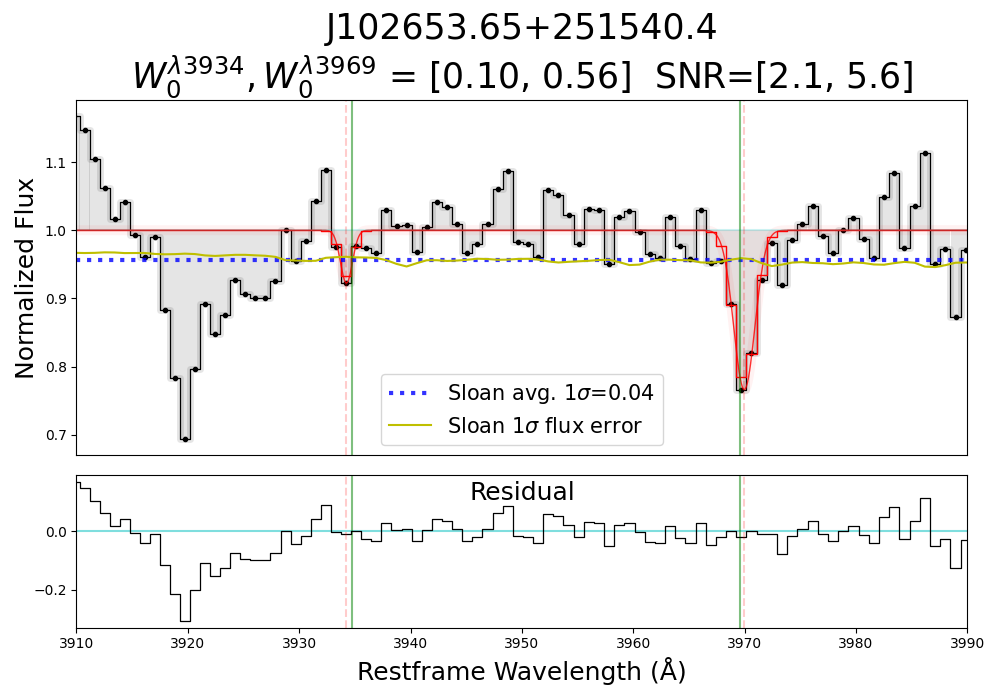}
    \includegraphics[width =.45\linewidth]{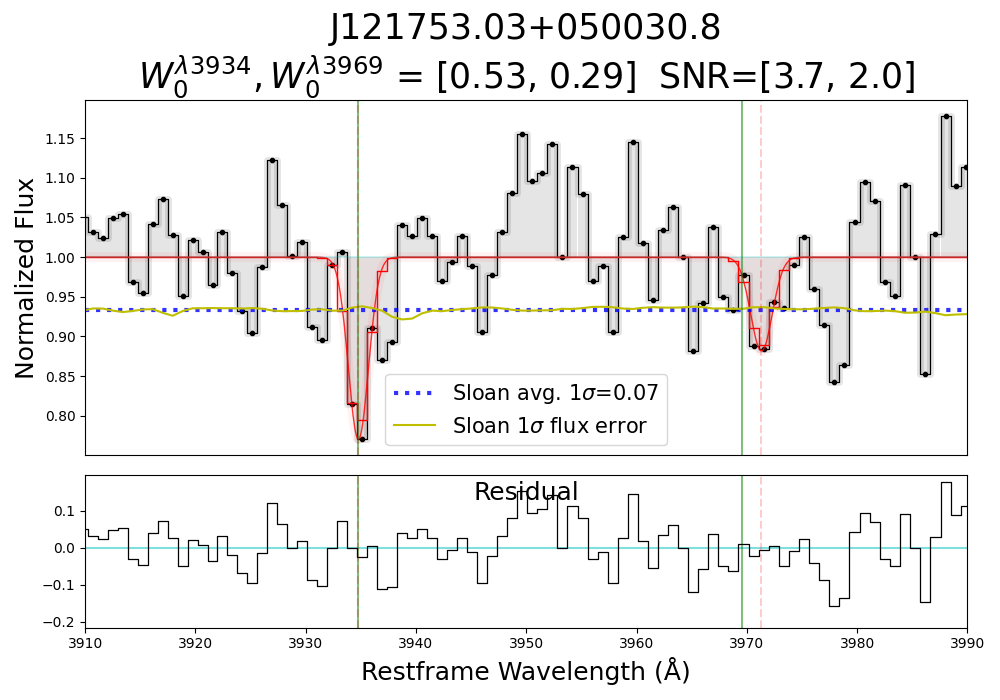}
    \caption{Detailed analyses of the six candidates in \citet{2014MNRAS.444.1747S}'s catalog that we found to be false Ca II absorbers. These candidates were not flagged by our detection program because they fall below our SNR criteria. Visual inspection showed them to have insignificant Ca II absorption. For each plot, the top panel contains the quasar spectrum and the line fittings while the bottom panel is the residual after the fit is removed. The vertical green line represents the Ca II absorber line's wavelength while the red dotted line denotes the Gaussian profile's center. The solid red line shows the Gaussian profile fitting both Ca II lines. The blue dashed line marks the average SDSS $1\sigma$ error and the yellow line shows the SDSS $1\sigma$ erorr at each wavelength.}
    \label{fig:SarFP}
\end{figure*}

228 of the 241 absorbers from \citet{2014MNRAS.444.1747S}'s catalog were flagged by our detection program. Thus, we were able to recover 95\% of the Ca II absorbers in \citet{2014MNRAS.444.1747S} catalog. The 13 undetected absorbers were measured by our detection program to be below our SNR thresholds. Visual inspection and further analysis of these undetected absorbers prove these candidates from Sardane et al.'s catalog to \textbf{not} be true Ca II absorbers. These false detections are presented in Table \ref{tab:Sardane_fp}.  Detailed plots of the first six false positive Ca II absorbers from \citet{2014MNRAS.444.1747S} catalog are shown in Figure \ref{fig:SarFP}. Based on the true Ca II absorbers from \citet{2014MNRAS.444.1747S}'s catalog that are present in the Mg II catalog we search, our detection program had a 100\% recovery rate.

The 228 Ca II absorbers in Sardane's catalog confirmed by our program were composed of 113 strong (56\% of all the strong absorbers in \citealt{2014MNRAS.444.1747S}'s catalog) and 115 weak absorbers (49\% of all the weak absorbers in \citealt{2014MNRAS.444.1747S}'s catalog). This consistent equal proportion of recovered weak and strong absorbers further demonstrates that our method of using Mg II absorbers as a probe to search for Ca II absorbers is not inherently biased towards finding weak or strong Ca II absorbers.

\begin{table}
\centering
\caption{\label{tab:Sardane_fp} Indentifying information for the false positive candidates we found in \citet{2014MNRAS.444.1747S}'s catalog.}
\begin{tabular}[t]{ccccc} 
\toprule
{Quasar} & {RA} & {Dec} & {$z_{\text{em}}$} & {$z_{\text{abs}}$}\\
\midrule
{J005355.15-000309.3}&{13.48}&{-0.05}&{1.70}&{1.25}\\
{J082918.09+113341.7}&{127.33}&{11.56}&{1.26}&{0.74}\\
{J094806.59+045811.7} & { 147.03 }&{ 4.97 }&{ 1.74 }&{ 0.90 }\\
{J100000.85+514416.6} & { 150.00 }&{ 51.74 }&{ 1.24 }&{ 0.91 }\\
{J102653.65+251540.4} & { 156.72 }&{ 25.26 }&{ 3.16 }&{ 0.65 }\\
{J121753.03+050030.8} & { 184.47 }&{ 5.01 }&{ 0.63 }&{ 0.54 }\\
{J124722.49+342727.0} & { 191.84 }&{ 34.46 }&{ 2.49 }&{ 1.04 }\\
{J140134.90+533714.2} & { 210.4 }&{ 53.62 }&{ 1.97 }&{ 0.83 }\\
{J144047.39+093809.0} & { 220.2 }&{ 9.64 }&{ 1.20 }&{ 0.90 }\\
{J152740.66+063218.5} & { 231.92 }&{ 6.54 }&{ 2.92 }&{ 0.45 }\\
{J155948.17+065727.6} & { 239.95 }&{ 6.96 }&{ 2.90 }&{ 0.53 }\\
{J160932.94+462613.3} & { 242.39 }&{ 46.44 }&{ 2.37 }&{ 0.97 }\\
{J213623.52-003410.9} & { 324.1 }&{ -0.57 }&{ 2.22 }&{ 1.22 }\\
\bottomrule
\end{tabular}
\end{table}

\subsection{Verifying Ca II absorbers}\label{sssec:verification}
 In verifying our flagged candidates as Ca II absorbers, we went through a three step process. First, we manually went through the over 1050 flagged candidates for review. Visual inspection of the fitted features was performed to make sure they were of the correct Ca II $\lambda \lambda$ 3934, 3969 features, and not of a different absorption line nearby or of noise. Emission lines were also taken into consideration during this step. We also reviewed the performed continuum fit during this visual stage. We note that a majority of the false detections were due to misfitted features, commonly due to the lack of Ca II absorption, meaning there was no Ca II absorption line feature to actually fit. Visual inspection brought the possible candidates down to over 700 candidates that we refit for the second stage of verification.
 
\begin{figure}
    \includegraphics[width =\linewidth]{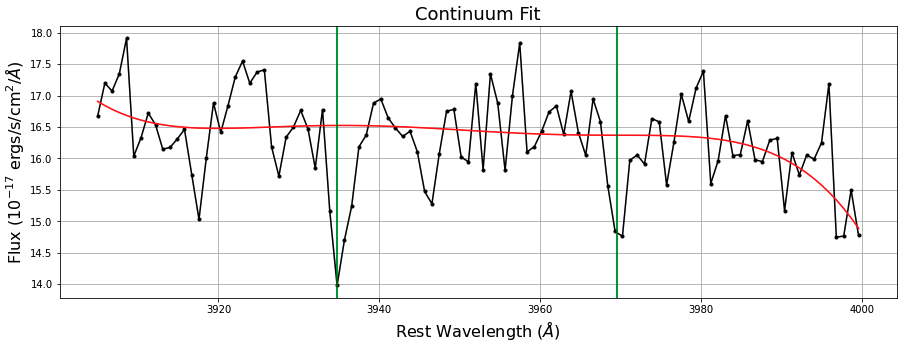}
    \caption{A continuum fitting for a candidate Ca II spectrum redshifted to restframe. For this fitting, a seventh degree polynomial was used. The normalized spectrum resulting from this continuum fit and its Gaussian profile fitting is shown in Figure \ref{fig:Norm_ex}}
    \label{fig:fig2}
\end{figure}

\begin{figure}
    \includegraphics[width =\linewidth]{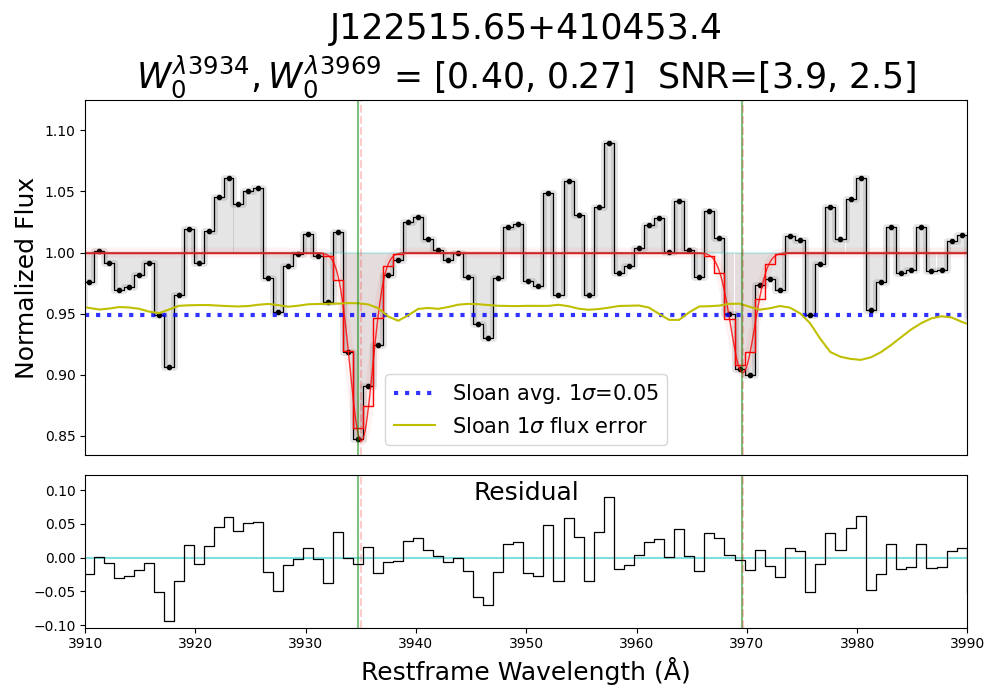}
    \caption{The normalized spectrum of Ca II absorption lines fitted with Gaussian profiles (solid red line) for equivalent width measurements. The top panel shows the quasar spectrum and the line fittings while the bottom panel is the residual after the fit is removed. The blue dashed line marks the average SDSS $1\sigma$ error while the yellow line shows the SDSS $1\sigma$ erorr at each wavelength.}
    \label{fig:Norm_ex}
\end{figure}

 In order to rule on candidates with borderline SNR detection and/or candidates facing visual scrutiny that may originate from the shift from observed to rest frame, we refit the candidates for more accurate $W_0$ measurements. With the raw flux, error, and corresponding wavelength data from SDSS, the spectra were recreated by interpolating between the points. Extreme points of the spectra (such as points of strong absorption or emissions) were rejected (filtered) for the continuum fitting. We then completed manual continuum fittings by choosing the polynomial order and level of rejection for the best possible continuum fit. This continuum is then used as a baseline to normalize the spectra. The normalized flux is calculated by dividing the smoothed flux by the continuum fit (normalized error is calculated similarly). Gaussian profiles were then fit to model the spectral line profiles for each absorber candidate. An example of a continuum fit and its corresponding normalized spectrum with Gaussian profiles is shown in Figure \ref{fig:fig2} and Figure \ref{fig:Norm_ex} respectively.
 
 To measure the strength of each absorption line, a model fitting the absorption line profile must be solved for the depth, $\sigma$, FWHM, and amplitude of the absorption line. For candidates in the weakest $W^{\lambda3934}_0$ regime, the maximum depth was constrained to the lowest point in a 5 \AA\ window, centered at the modeled line's wavelength to avoid overfitting of the line depth. These solved parameters were then used to calculate the $W_0$ of the target line. Both $W_0$ and its error were calculated with the formula described in the prior subsection. Gaussian profiles for the Ca II $\lambda 3934$ and $\lambda 3969$ absorption lines were solved independently. 
 
 For the third verification stage, Doublet Ratios (DR) of $W^{\lambda3934}_0/W^{\lambda3969}_0$ were observed as an indication of absorber's saturation. The Ca II $\lambda 3934$ line has two times the oscillator strength of the Ca II $\lambda 3969$ line \citep{1991ApJS...77..119M}. By keeping our catalog to $1-\sigma_{DR} \leq W^{\lambda3934}_0/W^{\lambda3969}_0 \leq 2+\sigma_{DR}$, where $\sigma_{DR}$ is the standard deviation of the DR measurements, we present systems between completely unsaturated and completely saturated extremes. Our three step verification process led to discovering 393 Ca II absorbers from SDSS DR7 and DR12. 
 
However, as described at the end of section \ref{sssec:discovery}, 228 of these Ca II absorber candidates, independently discovered by our detection program, were also in \citet{2014MNRAS.444.1747S}'s catalog. This brings our number of new Ca II absorber discoveries to 165.

These 393 Ca II absorbers were identified from $\sim$43,000 searched quasar spectra with Mg II absorbers, indicating only $\sim$0.9\% of Mg II absorbers are associated with detectable Ca II absorbers, hence, Ca II absorbers are indeed very rare. This rarity of Ca II absorbers is also consistent with \citet{2014MNRAS.444.1747S}'s findings, in which they identified 435\footnote{We note that we present 13 of \citet{2014MNRAS.444.1747S}'s absorbers as false detections in \ref{sssec:discovery}.} Ca II absorbers from $\sim$94,000 searched quasar spectra.
 
On the other hand, 228 of these Ca II absorber candidates, independently discovered by our detection program, were also in \citet{2014MNRAS.444.1747S}'s catalog, as described at the end of \ref{sssec:discovery}. This brings our number of new Ca II absorber discoveries to 165.

\subsection{Spectra stacking to measure element abundance}
To measure the relative abundances of $\alpha$ and Fe-peak elements, we need to derive column densities of different species from their measured equivalent widths, $W_0$ of their unsaturated absorption lines. However, as many of these absorption lines are too weak to be measured from individual quasar spectra, we combine different quasar spectra with Ca II absorbers to create  a composite spectrum from which the $W_0$ of many weak lines is measured. In other words, we stack the rest frame spectra of our absorbers together to create a median spectrum with reduced noise to have the most accurate mean measurement of different metal species of quasar absorption lines. While all the absorbers contain the Ca II $\lambda\lambda$3934, 3969 absorption lines to allow for an automatic stack of the entire catalog's spectra, other metal species require individual visual inspection of spectra. We used a three step process to decide whether an absorber's spectrum should be stacked for the metal line's final combine. We started with SDSS released quasar spectra and Doppler shifted it to its rest frame using the equation:
\[\lambda_{\text{rf}} = \lambda_{\text{obs}}/(1+z_{\text{abs}}),\] 
in which $\lambda_{\text{rf}}$ is the wavelength shifted to the rest frame and $\lambda_{\text{obs}}$ is the observed wavelength provided by SDSS. The first "test" in our process is whether the spectrum contains the full features after its shift back to the rest frame.  If not, the spectrum was rejected. If it does, the spectrum  was plotted for visual evaluation. With the same process described in the prior subsection, a continuum was fitted to the section of the spectrum within a window 90 \r{A} wide centered at the targeted line’s rest frame wavelength. The spectrum section was then normalized, and the normalized spectrum was plotted for visual inspection. Using these plots, we checked to see whether the spectrum has absorption features at the metal line's wavelength. 

Before going through each absorber's spectrum, one by one, we stack all the passed spectra and plot a figure that shows all the individual spectra against each other. We use this to find the metal line's average flux to help us in the next step of the evaluation. If it is known that the particular metal line we are working on is blended, we compare the line strengths of the lines blended together. We then use that comparison to figure out which condensed area of fluxes shown from the stacking of all passed ones was correct. We then examined each individual absorber's spectrum (if they passed the first examination) for the final two examinations, checking to see whether they 1) had a dip/flux past its average/surrounding noise at its wavelength, and 2) whether its flux at that line was around its  average flux predicted from other detected transitions of the same species. The expected average flux depended on the f-value of the measured line as strength of unsaturated lines follows f value for transitions of the same species.

The medians of the normalized fluxes from the absorbers kept were calculated and errors propagated for the creation of a final combine. Such combines allow for an accurate measure of the mean element abundances within the Ca II absorption-line systems. A Voigt profile is then fitted, with its solved parameters being used to calculate the $W_0$ of the target line. If the measured line’s absorption was not significant ($W_0/W_0(err)$ < $3\sigma$), the $W_0$ was presented by its upper limits.

\subsection{Search for 2175 \r{A} Dust Bump}\label{sssec:2DASearch}
We probed Ca II absorbers to search for whether their quasar spectra contained the 2175 \r{A} dust bump feature. The dust bump is only observable in absorption-line systems with $z_{abs} \geq 0.8$, so we only searched systems meeting this criteria. This amounted to 100 Ca II absorbers searched for the 2175 \r{A} dust bump.

To search for 2175 \r{A} dust absorbers (2DAs), we needed to solve for each candidate's extinction curve. First, the candidate system’s spectrum was dereddened by using \citet*{1998ApJ...500..525S}'s dust map (through the \texttt{sfdmap} Python module) to correct the Galactic extinction. A composite spectrum was then constructed by median combing all spectra in the SDSS released data as a non-absorber template to measure the color excess at the spectrum’s wavelengths (E($\lambda$-V); \citealt{2011ApJ...732..110J}). This was calculated by dividing reddened spectrum by the composite spectrum. Weights were then used to increase the priority of specific regions to be properly fit, like the region around the bump. Strong absorption lines and emission-like features were masked out so they would not influence the solved fits for extinction and the bump \citep{2011ApJ...732..110J, 2017MNRAS.472.2196M}.
For an accurate fit of the extinction and bump, three stages of fittings, increasing in complexity, were utilized to improve the parameters for the fitting curve \citep{2011ApJ...732..110J, 2015MNRAS.454.1751M}. A least-squares algorithm employing \texttt{LMFIT} for non-linear least-squares minimization was the first program used for curve fitting \citep{2015MNRAS.454.1751M}. The algorithm was first used to fit only the extinction to provide a rough estimate for the extinction curve parameters that are then used as initial estimations when the algorithm was used to solve for both the extinction and bump. 

We used the FM parameterized extinction curve to fit our extinction and bump as we follow the form $A(\lambda) = c_1+c_2x+c_3D(x,x_0,y)$ \citep{1990ApJS...72..163F, 2007ApJ...663..320F, 2003ApJ...594..279G, 2018MNRAS.474.4870M}. The extinction curve includes a Drude profile to measure the strength of the 2175 \AA\ dust bump. The Drude profile was calculated $D(x,x_0,y) = \frac{x^2}{(x^2-x_0^2)^2+x^2\gamma^2}$, where $x=\lambda ^{-1}$ in units of $\mu m^{-1}$. Our fitting models solved for five parameters: $c\textsubscript{1}$ (intercept), $c\textsubscript{2}$ (slope), c\textsubscript{3} (bump’s amplitude), $x\textsubscript{0}$ (position of bump peak), and $\gamma$ (FWHM of the Drude profile) in units of $\mu m^{-1}$. The strength of the bump is measured A\textsubscript{bump}$=\pi c_3/(2\gamma)$ in $\mu m^{-1}$. This represents the area above the extinction curve but below the linear component \citep{2017MNRAS.472.2196M}.

The third stage of fitting employed a two-stage Markov Chain Monte Carlo (MCMC) sampling approach. In a two-step process similar to the one taken in \citet{2017MNRAS.472.2196M}, we first fixed $\gamma$ and $x_0$ to the MW average values \citep{2007ApJ...663..320F} and solved for $c_1, c_2,$ and $c_3$. Then we fixed the $c_1, c_2,$ and $c_3$ values to those found in step one and solved for $\gamma$ and $x_0$. Doing this allows for better fittings as there are only two or three free variables during each MCMC sampler run, which decreases the initial variability in generated models. Leaving all parameters unconstrained ultimately results in an overly large gamma value because the MCMC sampler is trying to fit the extinction curve over the entire wavelength range while all variables are free.

\begin{figure}
    \centering
    \includegraphics[width =\linewidth]{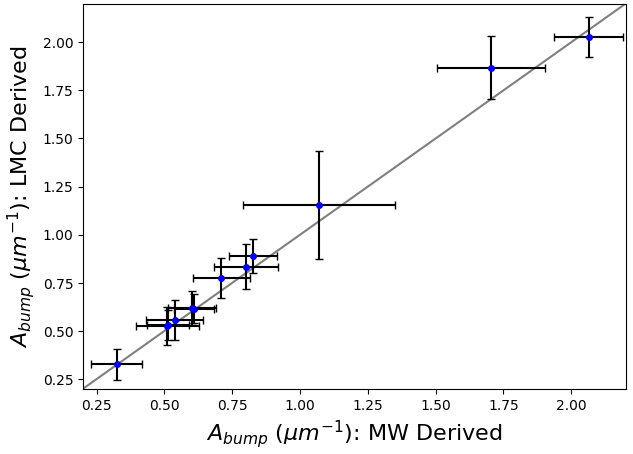}
    \caption{Bump strength derived from average MW $\gamma$ and $x_0$ values compared to bump strength derived from average LMC values. The error bars are errors propagated from the MCMC statistical uncertainties. The plot shows that there is little difference in MW and LMC derived measurements ($\sim4\%$).}
    \label{fig:figmwlmc}
\end{figure}

As 2175 \AA\ dust bumps are observed in both both MW and LMC sight lines, we evaluated how setting the initial fixed value of $\gamma$ and $x_0$ to either MW ($\gamma=0.89$ and $x_0=4.59$; \citealt{2007ApJ...663..320F}) or LMC\footnote{While \citep{2003ApJ...594..279G} reports different average FM parameter values for the LMC average sample and the LMC LMC2 Supershell Sample, the average $\gamma$ and $x_0$ values from the LMC average sample are within $1\sigma$ of those parameter values from the LMC LMC2 Supershell Sample.} ($\gamma=0.93$ and $x_0=4.58$; \citealt{2003ApJ...594..279G}) average values affect the final parameter values and 2175 \AA\ dust bump strength. Calculated dust bump strengths derived from average MW and LMC values are compared in Figure \ref{fig:figmwlmc}.  We observe generally wider LMC derived bump fits and slightly greater bumps strengths compared to those derived from MW values. 

However, as there is only a $\sim4\%$ difference between MW and LMC derived bump strengths, we decided to complete our initial discovery process using average MW values (following \citealt{2018MNRAS.474.4870M}'s precedent of doing so).

% \citet{2015MNRAS.452.3192S} found that  both the LMC and SMC dust laws provided good fits to the observable extinction of their quasar spectra with Ca II absorbers.  However,  since the 2175 \r{A} dust bump feature is only observed in the LMC extinction curve, we use }}the LMC $\gamma$ parameter. Setting $\gamma$ to this static value provided a starting point, one likely near the true fit’s value, which allowed the fitting model to accurately solve for $c\textsubscript{1}$, the intercept of the parameterized extinction curve. 

The MCMC solver (from the \texttt{emcee} package from \citet{2013PASP..125..306F}) runs 3,000 steps (simulations by each walker) but we discarded (“burn in”) the first 500 to only sample distributions in a high probability region in our result. Log likelihoods were used for evaluation of each step’s model data versus the current best model which allows for subsequent steps to be towards regions of high likeliness between generated models. A successful MCMC sampler would produce a normal posterior probability distribution of generated model likelihoods. 

The posterior distributions, by parameter, were visualized with a corner plot that shows all the 1D and 2D projections along with the parameters, valued at the greatest likelihood. We evaluated the corner plot by examining the distributions. The $c\textsubscript{1}$ parameter was then set as a static parameter with the solved value. If the $x\textsubscript{0}$ parameter’s distribution was skewed, then we also limited the possible range of values for the parameter. The MCMC solver was then run again with initial guesses of the first MCMC’s solved parameters and the ones modified ($c\textsubscript{1}$, possibly $x\textsubscript{0}$). The main goal of this process was to derive the best possible fit for $x\textsubscript{0}$ and $\gamma$. 

The resulting parameters from the second stage of MCMC sampling were used to fit the extinction curve and evaluate the candidacy of the absorber as a dust absorber. From the fitted extinction curve, we also estimated A\textsubscript{V}, the total amount of extinction at the V band ($\lambda\textsubscript{Obs}=5500$ \r{A}). The solved parameters were then used to measure the area of the Drude profile ($A\textsubscript{bump}$) to evaluate the strength of the 2175 \r{A} dust bump and whether the bump was significant (bump significance level $\geq 3\sigma$ limits). If the bump was significant, then the Ca II absorber was also classified as a 2DA. 
 
To account for the difference in bump strength derived from different initial values of bump width $\gamma$, we refit each 2DA using average LMC values for our MCMC sampler to verify our discoveries. We only considered Ca II absorbers with consistently significant bump strengths ($\geq 3\sigma$ limits for both MW and LMC derived bump strengths) as real 2DAs. Reported FM parameters for each 2DA in results are averaged between the MW and LMC derived values. In total, we discovered 12 2DAs in our catalog.

\section{\fontsize{12}{15}\selectfont Results}
\subsection{Ca II Catalog Statistics}
Our Ca II catalog is composed of 165 new Ca II absorbers that were discovered with a $3\sigma$ and $2.5\sigma$ threshold for $\lambda 3934$ and $\lambda 3969$ line, respectively.  Target information, equivalent widths and redshift information for the first 10 Ca II absorbers can be found in Table ~\ref{tab:Table 1}.

\begin{table*}
\centering
\caption{\label{tab:Table 1} The Ca II catalog. The entire catalog is available in its entirety online.}
\begin{tabular}[t]{ccccc} 
\toprule
{Quasar} & {$z_{\text{em}}$} & {$z_{\text{abs}}$} & {$W_0^{\lambda3934}$} & {$W_0^{\lambda3969}$} \\
{} & {} & {} & {(\AA)}&{(\AA)}\\
\midrule
{J093256.80+074212.2}&{1.00}&{0.72}&{$0.53\pm0.09$}&	{$0.30\pm0.10$}\\
{J102837.01-010027.4} & { 1.53 } & { 0.63 } & {$0.32\pm0.09$} & {$0.27\pm0.10$}\\
{J112835.39+414327.9} & { 1.00 } & { 0.52 } & {$0.30\pm0.10$} & {$0.13\pm0.05$}\\
{J122515.65+410453.4} & { 1.16 } & { 0.91 } & {$0.40\pm0.10$} & {$0.27\pm0.10$}\\
{J123332.96+080018.0} & { 1.52 } & { 1.14 } & {$ 0.83\pm0.23 $} & {$ 0.41\pm0.16 $}\\
{J143120.53+395241.5} & { 1.22 } & { 0.60 } & {$ 0.27\pm0.09 $} & {$ 0.20\pm0.08 $}\\
{J015249.76+002314.5} & { 0.59 } & { 0.48 } & {$ 0.34\pm0.11 $} & {$ 0.29\pm0.09 $}\\
{J075123.60+084248.8} & { 1.55 } & { 0.54 } & {$ 0.85\pm0.09 $} & {$ 0.50\pm0.09 $}\\
{J110736.67+000329.4} & { 1.73 } & { 0.95 } & {$ 0.48\pm0.11 $} & {$ 0.38\pm0.15 $}\\
{J110931.16+353025.8} & { 1.00 } & { 0.57 } & {$ 0.60\pm0.08 $} & {$ 0.41\pm0.08 $}\\
\bottomrule
\end{tabular}
\end{table*}

\begin{figure}
    \centering
    \includegraphics[width =\linewidth]{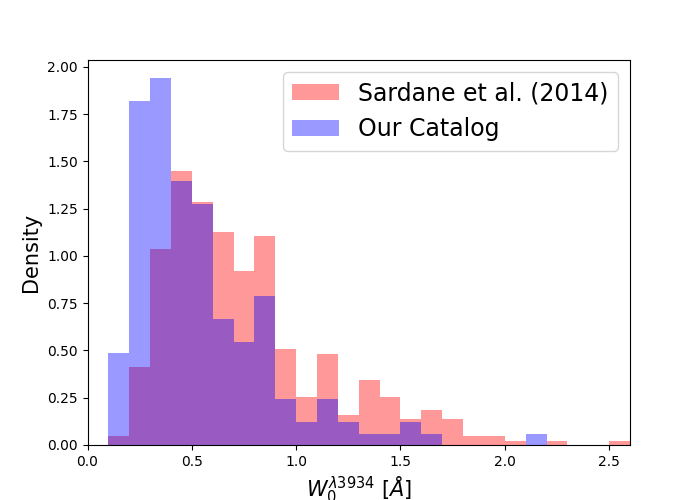}
    \caption{The Distribution of the equivalent widths of $\text{W}^{\lambda3934}_{0}$. It is evident that there are more candidates with $\text{W}^{\lambda3934}_{0}<0.7$ \AA, which we consider as the weak absorber subpopulation compared to strong absorber subpopulation. The figure compares the density of our catalog's $\text{W}^{\lambda3934}_{0}$ distribution to that of Sardane et al's (2014).}
    \label{fig:figew}
\end{figure}

The strongest absorber that we found has $\text{W}^{\lambda3934}_{0}$ = 2.10 \AA, and our weakest absorber has $\text{W}^{\lambda3934}_{0}$ = 0.13 \AA.  This distribution also has a mean of $\overline{\text{W}^{\lambda3934}_{0}}=0.54$ \AA.  The standard deviation is $\sigma$ = 0.33  \AA. Figure ~\ref{fig:figew} shows our catalog's distribution of $\text{W}^{\lambda3934}_{0}$ compared to that of \citet{2014MNRAS.444.1747S}'s. Our catalog has a greater proportion of absorbers with $\text{W}^{\lambda3934}_{0}<0.7$ \AA\ (i.e., "weak" absorbers) than \citet{2014MNRAS.444.1747S}. Our catalog is also skewed towards the lower $\text{W}^{\lambda3934}_{0}$ regime. The concept of two distinct populations for Ca II quasar absorbers, as proposed by \citet{2014MNRAS.444.1747S}, is also supported by our bimodal distributions of the DR.

\begin{figure}
    \centering
    \includegraphics[width =\linewidth]{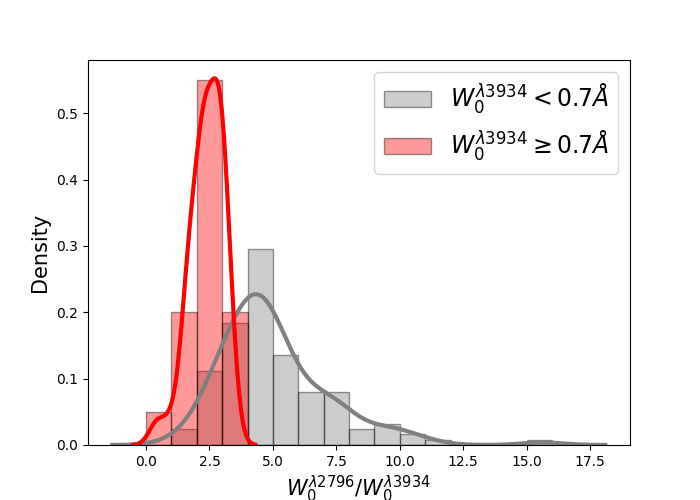}
    \caption{Distribution of equivalent width ratio of the Mg II $\lambda2796$ absorption line to the Ca II $\lambda3934$ absorption line.  In this figure, it is apparent that there are two populations. The grey distribution shows density of the "weak" Ca II absorbers' (absorbers with $\text{W}^{\lambda3934}_{0}$<0.7 \AA) $\text{W}^{\lambda2796}_{0}/\text{W}^{\lambda3934}_{0}$ ratio and the grey line is its corresponding distribution curve. The red distribution shows density the "strong" Ca II absorbers' (absorbers with $\text{W}^{\lambda3934}_{0}$>0.7 \AA) $\text{W}^{\lambda2796}_{0}/\text{W}^{\lambda3934}_{0}$ ratio and the red line is its corresponding distribution curve. The distribution curves are calculated using \texttt{seaborn's} kernel density estimation function..}
    \label{fig:figtwopop}
\end{figure}

We separate our catalog into two subpopulations of weak ($\text{W}^{\lambda3934}_{0} < 0.7$ \AA) and strong ($\text{W}^{\lambda3934}_{0}\geq0.7$ \AA) absorbers and plot the distribution of Mg II $\text{W}^{\lambda2796}_{0}$ divided by Ca II $\text{W}^{\lambda3934}_{0}$ in Figure ~\ref{fig:figtwopop}. The result is a bimodal distribution, more clearly shown with the distribution curves, calculated using \texttt{seaborn's} kernel density estimation function \citep{Waskom2021}, in Figure ~\ref{fig:figtwopop}. The existence of a bimodal distribution from the two subpopulations in our catalog supports the bimodal distribution that was produced by \citet{2014MNRAS.444.1747S}. Our catalog contains 125 weak and 40 strong absorbers.

Given that \citet{2014MNRAS.444.1747S} searched DR7 and DR9, both of which we also searched (given that DR12 contains all DR9 targets), we believe our small proportion of strong absorbers found is due to \citet{2014MNRAS.444.1747S} already discovering a majority of them. 

The end of section \ref{sssec:discovery} discusses how our detection method is not inherently biased towards discovering weaker Ca II absorbers as it discovered about the same amount of strong and weak absorbers from \citet{2014MNRAS.444.1747S}'s catalog. As described in \ref{sssec:verification}, our detection program actually independently found 393 real Ca II absorbers, but 228 overlapped with those found in \citet{2014MNRAS.444.1747S}'s catalog. Of the original found sample of 393 Ca II absorbers, from DR7, we discovered 13 weak and 6 strong absorbers in addition to \citet{2014MNRAS.444.1747S}'s 43 weak and 38 strong absorbers and from DR12, we discovered 112 weak and 34 strong absorbers in addition to \citet{2014MNRAS.444.1747S}'s 72 weak and 75 strong absorbers. While the proportion of new strong absorbers discovered by our detection program is small compared to that of new weak absorbers discovered, the proportion of strong absorbers discovered that overlap with \citet{2014MNRAS.444.1747S}'s catalog is about the same as that of the overlapping weak absorbers discovered (both $\sim$50\%).

Our ability to detect about 2x more weak Ca II absorbers missed by \citet{2014MNRAS.444.1747S} than strong ones in DR7 likely implies that our detection program is more sensitive, allowing it to detect smaller features more efficiently. Further, our detection program finding only 40 new strong Ca II absorbers compared to 125 new weak absorbers could be a reflection of this increased ability to detect weak absorbers. This implies that strong Ca II absorbers are rarer than weak Ca II absorbers.


\begin{figure}
    \centering
    \includegraphics[width =\linewidth]{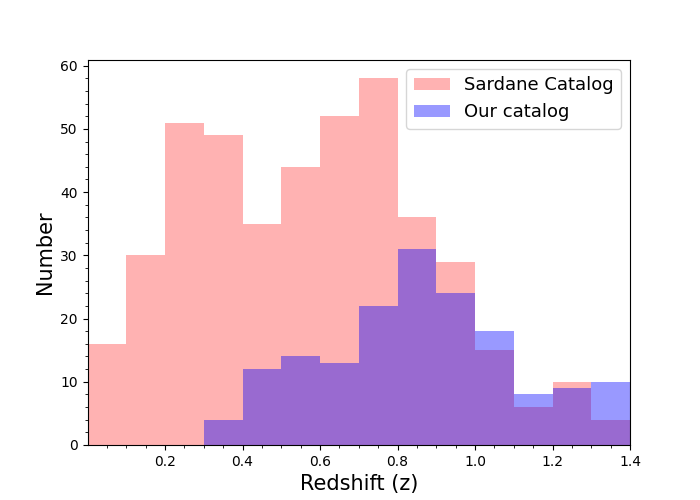}
    \caption{The distribution of absorber redshifts for the our catalog of Ca II absorbers compared to that of \citet{2014MNRAS.444.1747S}.  Our mean redshift was 0.91, and there is a standard deviation of $\sigma$ = 0.40.  It is apparent that our distribution is more skewed towards the higher redshift regime than \citep{2014MNRAS.444.1747S}'s distribution.}
    \label{fig:figzabs}
\end{figure}

In our catalog, all of the absorbers have redshifts 0.35<z\textsubscript{abs}<1.4. The lower redshift limit results from the use of Mg II absorbers in Ca II absorber detection. Figure \ref{fig:figzabs} shows the distribution of our catalog's z\textsubscript{abs} compared to \citet{2014MNRAS.444.1747S} catalog. Our catalog is skewed towards the higher reshift regime and our catalog contains a greater proportion of high redshift absorbers compared to \citet{2014MNRAS.444.1747S}'s. While both our catalog and \citet{2014MNRAS.444.1747S}'s contains 100 absorbers with $z\geq0.8$, that constitutes 61\% of our catalog compared to 23\% of \citet{2014MNRAS.444.1747S}'s catalog. The distribution features a mean redshift of <z\textsubscript{abs}> = 0.86 and a standard deviation $\sigma_{\text{abs}}$ = 0.25.  

\begin{figure}
    \includegraphics[width =.5\linewidth]{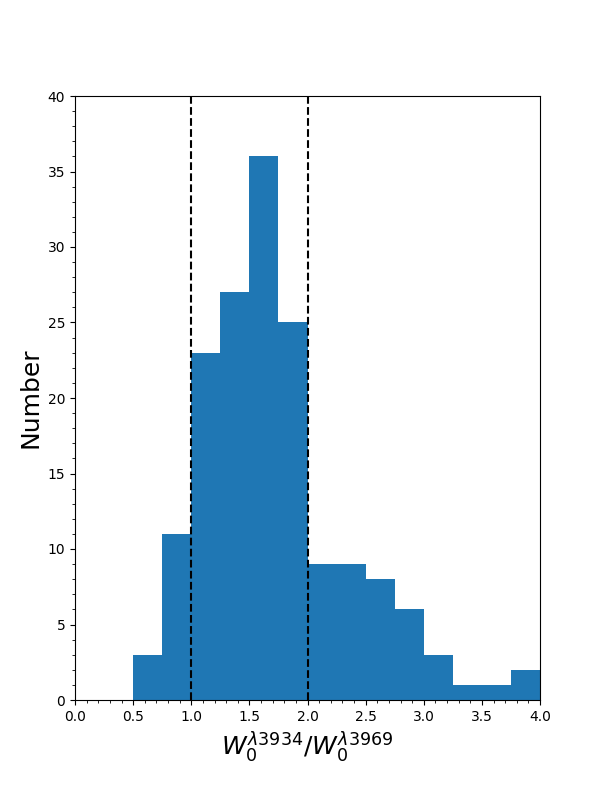}
    \includegraphics[width =.5\linewidth]{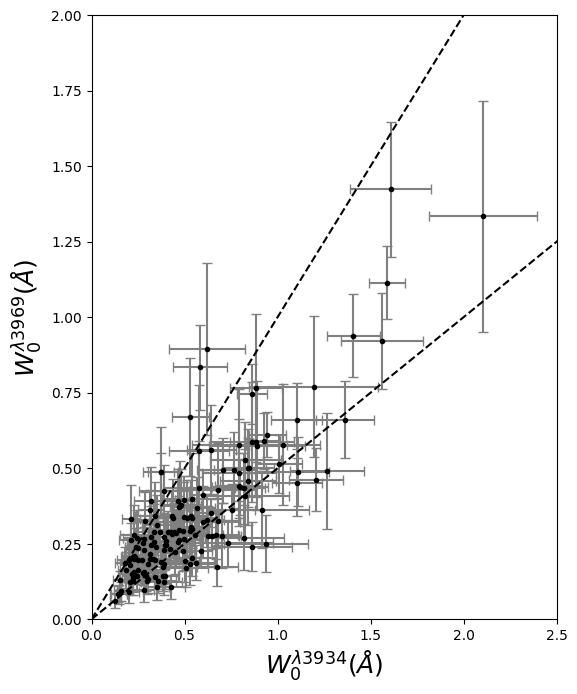}
    \caption{Left: The distribution of DRs ($\text{W}^{\lambda3934}_{0}$/$\text{W}^{\lambda3969}_{0}$). Right: $\text{W}^{\lambda3934}_{0}$ versus $\text{W}^{\lambda3969}_{0}$. For both figures, the dashed lines represent the limits for completely saturated and unsaturated systems at DR = 1.0 and DR = 2.0 respectively.}
    \label{fig:figdr}
\end{figure}

\begin{figure}
    \centering
    \includegraphics[width =\linewidth]{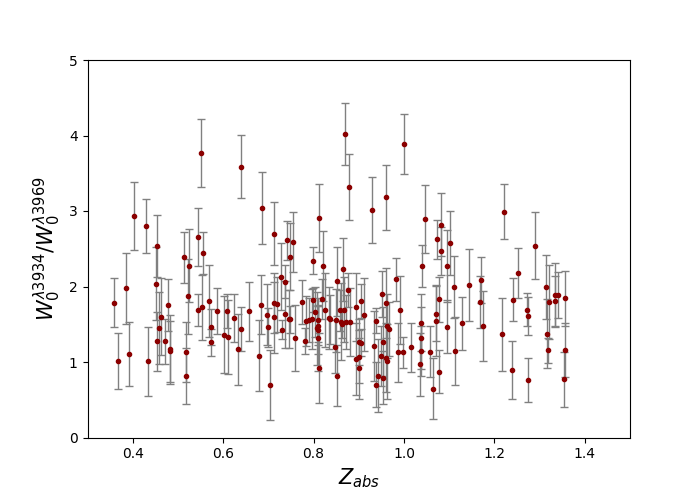}
    \caption{DR plotted as a function of redshift.  There does not appear to have any correlations.}
    \label{fig:figdrzabs}
\end{figure}

The DR of $\text{W}^{\lambda3934}_{0}$/$\text{W}^{\lambda3969}_{0}$ indicates the saturation of the observed system, with a DR range of around 1 to 2 for completely saturated to unsaturated systems, respectively. In our catalog, the mean of the DRs is 1.72, and the standard deviation is $\sigma$ = 0.65. Figure \ref{fig:figdr} shows the distribution of the DR.  In Figure \ref{fig:figdrzabs}, the DR is plotted as a function of redshift, showing no detectable evolution in the DR distribution.

\begin{figure}
    \centering
    \includegraphics[width =\linewidth]{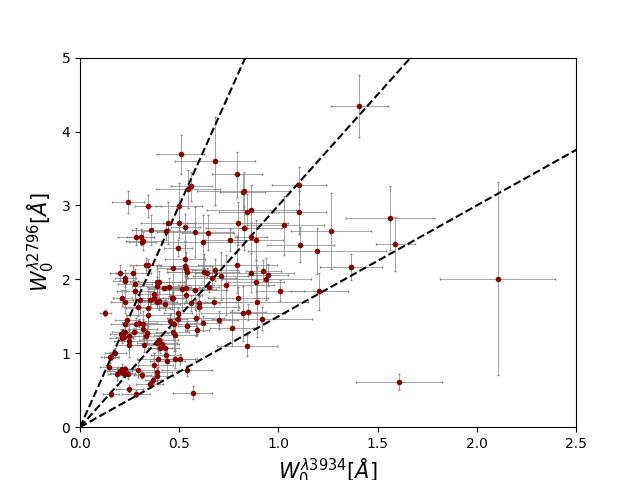}
    \caption{$\text{W}^{\lambda2796}_{0}$ versus $\text{W}^{\lambda3969}_{0}$ for all 165 Ca II systems in our catalog. The three dotted lines show $\text{W}^{\lambda2796}_{0} / \text{W}^{\lambda3969}_{0} = [1.5, 3, 6]$.}
    \label{fig:figmgca}
\end{figure}

As our catalog of Ca II absorbers was detected from a catalog of Mg II absorbers, we have all 165 Ca II-Mg II systems to use for comparisons. The K lines from both elemental doublets were plotted against each other and that is shown in Figure \ref{fig:figmgca}. There is a weak positive correlation, with R=0.42, between the strengths of the Mg II $\lambda$3934 and Ca II $\lambda$3934 lines.

\subsection{CaII $\lambda\lambda$3934, 3969 combined spectrum and equivalent widths}
\begin{figure}
    \centering
    \includegraphics[width =\linewidth]{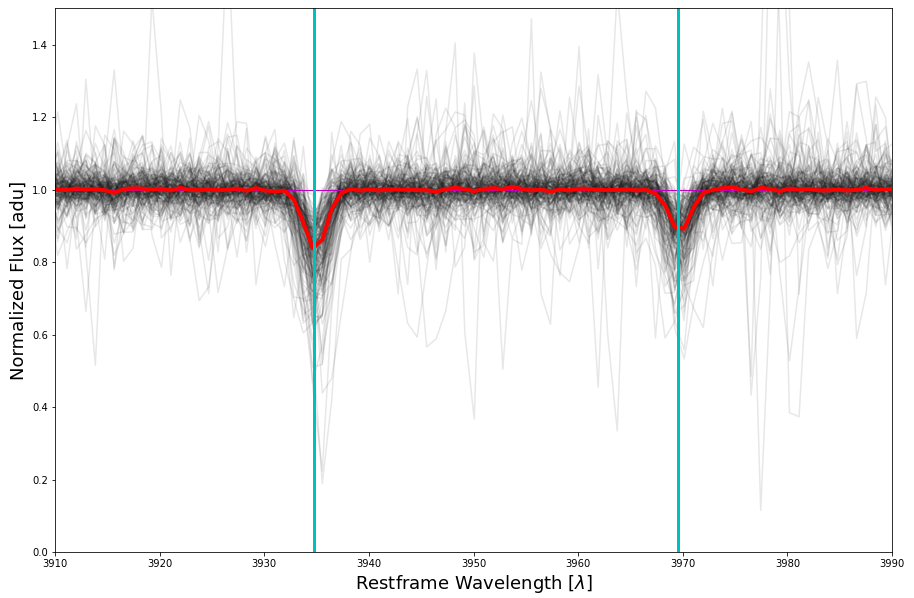}
    \caption{The composite spectrum for our catalog of CaII $\lambda\lambda$3934, 3969 absorption lines is shown in red. Individual spectra from our catalog are plotted in black behind the composite.}
    \label{fig:fig3}
\end{figure}

\begin{figure}
    \centering
    \includegraphics[width =\linewidth]{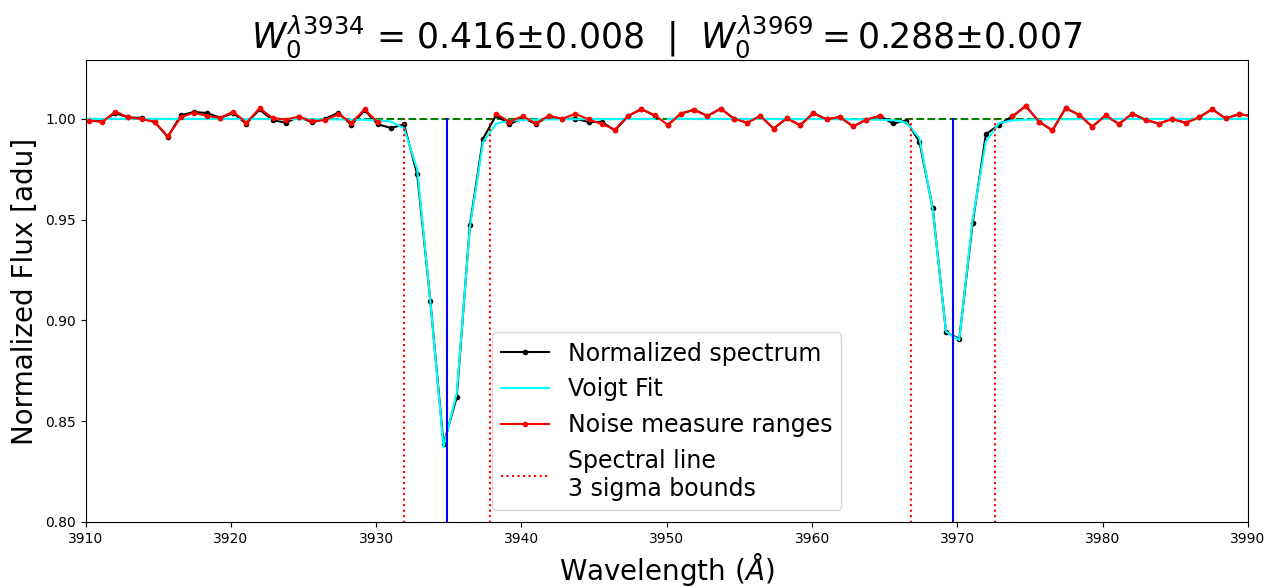}
    \caption{Voigt profiles fitted to the Ca II $\lambda\lambda3934, 3969$ lines for the catalog's composite spectrum are shown in cyan. $W_0$ error is calculated by multiplying the FWHMs of the Voigt profile by the rms of the flux from the noise measure ranges (in red).}
    \label{fig:fig5}
\end{figure}

\begin{figure*}
    %\centering 
    \advance\leftskip-1.7cm
    %[width =.9\paperwidth,scale=0.25]
    \includegraphics[width =\paperwidth]{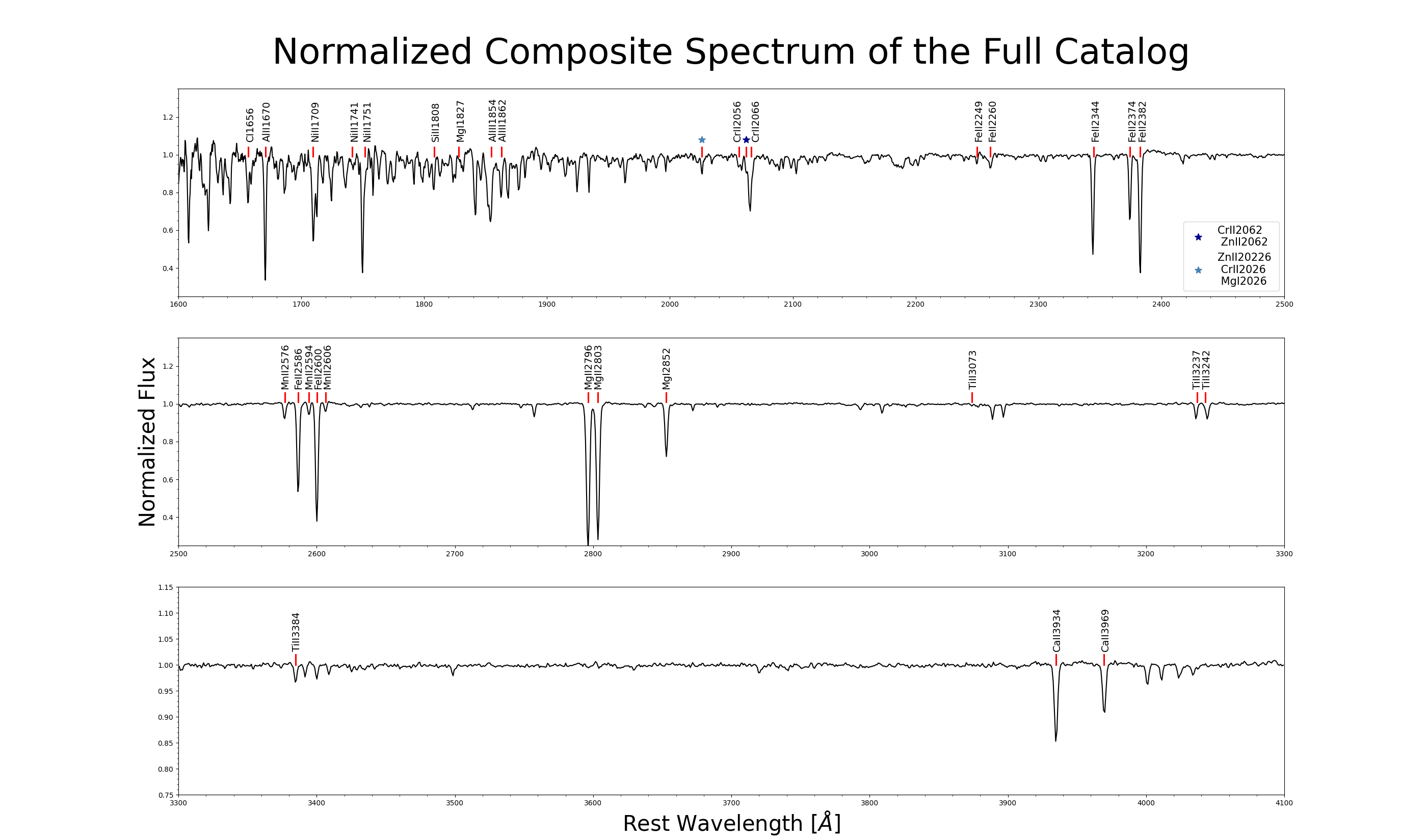}
    \caption{Normalized composite spectrum of the full catalog. Red lines mark the absorption lines that were later measured for element abundance measurements.}
    \label{fig:figfnc}
\end{figure*}

Using the spectra stacking method from Section 2.3 to improve SNR, we combined the spectra of CaII$\lambda\lambda$3934 and 3969 (respectively) in our catalog. The final combine for CaII$\lambda\lambda$3934, 3969 can be seen in Figure~\ref{fig:fig3}. These were fit with a Voigt function using a least squares fitting method. The resulting parameters from this fit was then used to measure the spectral line's $W_0$. Figure~\ref{fig:fig5} shows the Voigt fitiings and $W_0$s for CaII $\lambda\lambda$3934 and 3969. The normalized composite spectrum of the full catalog was formed to exhibit the mean abundance of absorption features within the Ca II absorber population. Some absorption lines, which are dominant species in different elements are difficult to identify in individual spectra due to weak transitions (such as Mn II, Cr II, Zn II, and Ni II), but they are clearly visible in the normalized composite spectrum for deriving mean column densities of these elements from their measured equivalent widths, $W_0$s. This is an advantage of analyzing the entire catalog through stacked spectra instead of focusing on several systems. The normalized composite spectrum constructed from all 165 Ca II absorbers is shown in Figure~\ref{fig:figfnc}. For our composite spectra, we used the rms of nearby regions to measure noise and multiplied it by the FWHMs of the Voigt profile to calculate $W_0$ error. To test the accuracy of our methodology, we also compared the measured errors to those derived from error propagation from spectrum stacking. For instance, the error of $W_0^{\lambda3934}$ from using error propagation was 0.0065 \AA\ while the error calculated from using nearby noise ranges was 0.0069 \AA, which are consistent with each other. For $W_0^{\lambda3969}$, its error from the error  propagation was 0.0066 \AA\ while its error derived from using nearby noise ranges was 0.0067 \AA. The difference between the two methods' results is negligible. We adopted the second method to derive $W_0$ errors for other species in the following subsection. 

\subsection{Measuring column densities}
To learn about the properties of Ca II quasar absorbers as a population, we derived element column densities to calculate the abundance of elements in Ca II absorption-line systems for comparison to the Milky Way’s interstellar media. First, the $W_0$s  of 40 different absorption lines (all marked in Figure \ref{fig:figfnc}) in our Ca II absorption-line systems were measured. The two distinct Ca II absorber populations were also taken into consideration. These measurements were obtained for the entire catalog and two subsamples. Absorbers were placed in two subsamples, based on the distinct population they fell into. The technique of spectra stacking for a final combine of spectra with targeted absorption lines and the use of Voigt profiles as described in the methods Section were used.

Once $W_0$ was determined for all 40 target lines, the Fe II lines were used to generate a curve of growth. The curve of growth, needed to determine which absorption lines in the system are saturated, is created by plotting $\text{log}_{10}(\text{N}f\lambda)$ against $\text{log}_{10}(\frac{\text{W}}{\lambda})$ of different species. The unsaturated lines fall on the linear part of the curve of growth while the saturated lines are in the curved section. The column densities of absorption lines that are not saturated were calculated with the equation $N \approx 1.13*10^{20} \frac{W_0}{f\lambda^2}$ cm$^{-2}$, where both ${W_0}$ and $\lambda$ are in units of \AA. These column densities were then used to evaluate the abundance of elements in the gas phase within the Ca II absorption line systems.
\begin{figure}
    \centering
    \includegraphics[width =\linewidth]{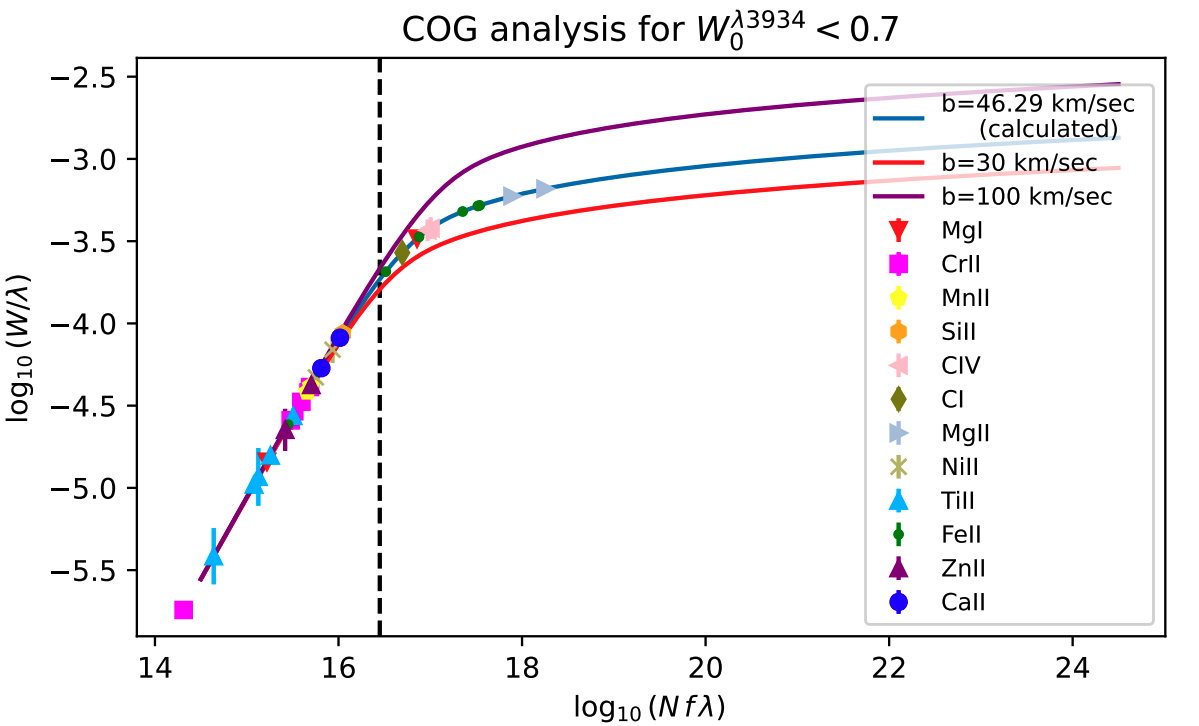}
    \includegraphics[width =\linewidth]{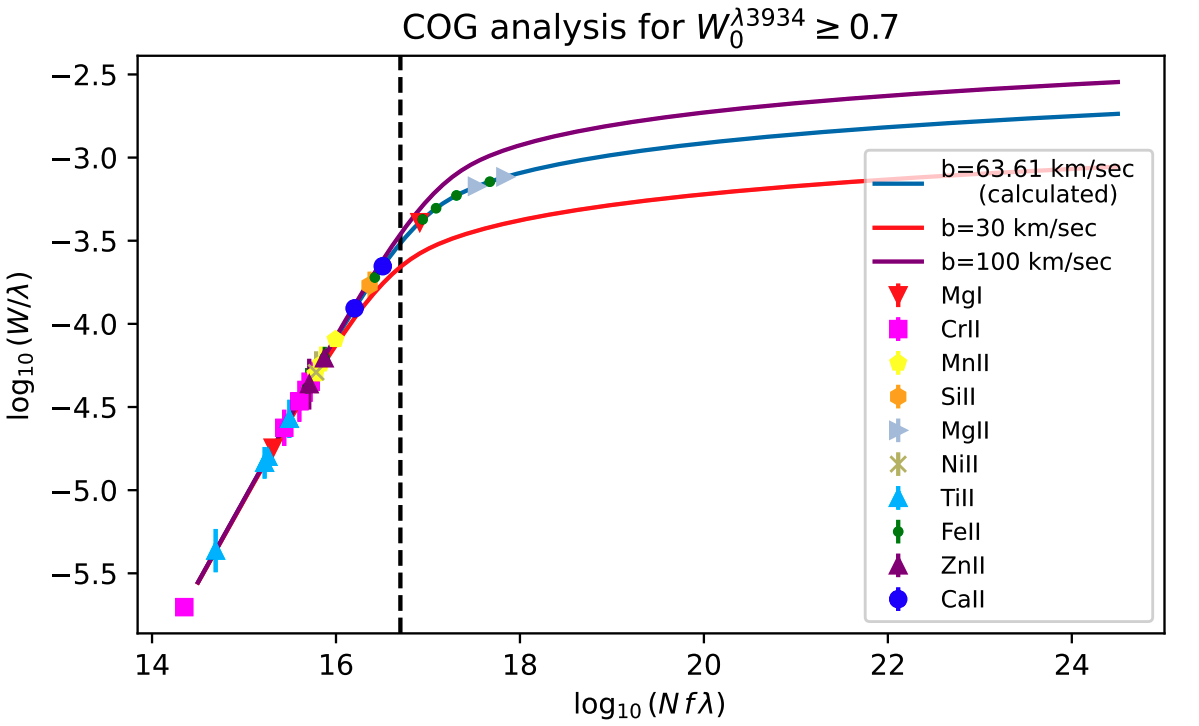}
    \caption{Curve of growth (COG) analyses for each subpopulation. The top figure analyzes the weak population and the bottom figure analyzes the strong population. The dotted line denotes the point of saturation. The strong population's COG does not include CI and CIV lines because the sample was too small to concretely measure  $W_0$.}
    \label{fig:figcog}
\end{figure}

Curve of growth analyses were performed separately for the strong and weak Ca II absorber populations and are shown in Figure~\ref{fig:figcog}. The Doppler parameter ($b$) from the curves of growth were also derived. Comparison of the strong and weak population’s Doppler parameters supports the evidence that they are distinct populations. The Doppler parameter of the strong population is significantly larger than that of the weak population as 63.6 > 46.3 km/s. This also indicates that the strong population features a greater component makeup than that of the weak population on average.

We use the calculated Doppler parameter as an indicator of the velocity dispersion of components because the velocity components cannot be resolved by the low-resolution of SDSS spectra. As the Doppler parameter is defined as $b=\sqrt{2}\sigma$ (where $\sigma$ is the velocity dispersion), the velocity dispersion can be derived from $b$  \citep{2011piim.book.....D}. Applying this to the solved Doppler parameters, we get $\sigma=32.7 km/s$ for the weak Ca II population and $\sigma=45.0 km/s$ for the strong Ca II population. SDSS DR7 has a resolving power (R) of 1850-2200 and an average spectral resolution ($\delta v=c/R$) of $\sim150$ km/s and SDSS DR12 has a resolving power (R) of 1500-2500 and an average spectral resolution ($\Delta v=c/R$) of $\sim150$ km/s \citep{2009ApJS..182..543A, 2015ApJS..219...12A}. Both the estimated velocity dispersions for the weak and strong Ca II absorber populations are more than 3 times less than the average spectral resolution of the SDSS spectra used. Further, SDSS DR7 cautions against using SDSS velocity dispersion measurements when estimates are <70 km/s and SDSS DR12 cautions against using measurements <100 km/s because they are below the resolution limit of the SDSS original spectrograph and BOSS spectrograph, respectively. Thus, we cannot resolve our line profiles with these velocity dispersion. Instead, we rely on the calculated Doppler parameter values from our generated curve of growth to intrinsically evaluate the FWHM of our line profiles (1.665$b$) and velocity dispersion. Doing so does not decrease the accuracy of the element abundances described in the next section because we only factor in measurements from unsaturated lines (located along linear portion of the curve of growth), which are not affected by Doppler broadening.

\subsection{Relative Element Abundances}
\begin{figure}
    \centering
    %\advance\leftskip-1cm
    \includegraphics[width =\linewidth]{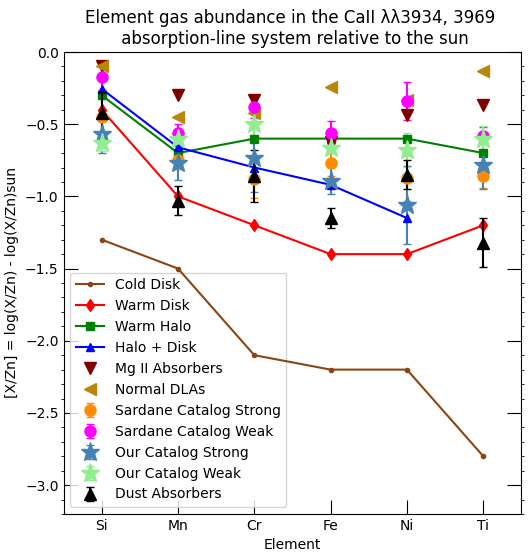}
    \caption{Elemental abundances relative to Zn for subpopulations $W^{\lambda3934}_0\geq0.7 $\AA\ (strong) and $W^{\lambda3934}_0<0.7 $\AA\ (weak) from our catalog and \citet{2014MNRAS.444.1747S}'s catalog. Element abundances from various Milky Way galactic components are plotted for comparison \citep{1996ARA&A..34..279S, 1999ApJS..124..465W}. Mg II absorber element abundances averaged from \citet{2006MNRAS.367..945Y}, dust absorber element abundances averaged from \citet{2018MNRAS.474.4870M}, and normal DLA element abundances from \citet{2016MNRAS.458.4074Q} are also plotted for comparison. Our catalog's strong subpopulation shows stronger similarities to the element depletion characteristics of the Halo + Disk component than Sardane et al.'s except Si. Our catalog's weak subpopulation also has stronger similarities to the the element depletion characteristics of the Warm Halo component than Sardane et al.'s except Si. More detailed comparisons are reported in the text. }
    \label{fig:figabun}
\end{figure}

Abundances are assessed relative to the Sun with Zn II as a metallicity indicator \citep*{1990ApJ...348...48P}. Relative abundances are defined by $[\frac{m}{Zn}] = \text{log}[\frac{N(m)}{N(Zn)}] - \text{log}[\frac{N(m_{\text{sun}})}{N(Zn_{\text{sun}})}]$ in which $m$ is the element for which the abundance is being measured. $N(m)$ is the mean column density of the element’s unsaturated absorption lines,  and solar values are from \citet{2009ARA&A..47..481A}. Relative abundances of the two populations were plotted in comparison to those of cold disc gas, warm disc gas, disc + halo gas, and warm halo gas components of the Milky Way’s interstellar media (compiled from \citet{1996ARA&A..34..279S, 1999ApJS..124..465W}). The result, shown in Figure~\ref{fig:figabun}, further affirms that Ca II absorbers are composed of two distinct populations. Elemental abundances are also reported in Table \ref{tab:Table 2}.

\begin{table}
\centering
\caption{\label{tab:Table 2} Element abundances relative to Zn for the two Ca II absorber subpopulations. These abundances are plotted in comparison to \citet{2014MNRAS.444.1747S}'s catalog and various Milky Way galactic components in \ref{fig:figabun}}.
\begin{tabular}[t]{ccc}
\toprule
 \multicolumn{3}{|c|}{[$X$/Zn]} \\
{X} & {$W^{\lambda3934}_0<0.7$\AA} & {$W^{\lambda3934}_0\geq0.7$\AA}\\
\midrule
{Si}&{$-0.63\pm0.06$}&{$-0.57\pm0.13$}\\
{Mn}&{$-0.60\pm0.04$}&{$-0.77\pm0.12$}\\
{Cr}&{$-0.50\pm0.11$}&{$-0.73\pm0.24$}\\
{Fe}&{$-0.67\pm0.03$}&{$-0.89\pm0.09$}\\
{Ni}&{$-0.68\pm0.12$}&{$-1.06\pm0.27$}\\
{Ti}&{$-0.61\pm0.09$}&{$-0.78\pm0.17$}\\
\bottomrule
\end{tabular}
\end{table}

The strong population’s element abundances are generally consistent with those of the Milky Way’s halo + disc components except Si. The consistency is clear, and our catalog’s Ca II absorbers prove a stronger correlation with the halo + disc components than in the results presented by \citet{2015MNRAS.452.3192S}. The weak population’s element abundances are generally consistent with those of the Milky Way’s warm halo components except Si. As different components are consistent with the two Ca II absorber populations, their environments will also be different. Our results indicate that strong Ca II absorbers are highly likely associated with disc-like environments. This conclusion is supported by the dustiness of the strong population (see Sections \ref{sssec:2DA}, \ref{sssec:reddening}) as galactic disks generally feature ongoing star formation and dust is a critical component in star formation. Consistency with warm halo components indicate that weak Ca II absorbers are highly likely associated with galactic halo environments. Destruction of dust is more severe in the halo’s environment than in the disc’s and this corresponds with a lower dust content in weak Ca II absorption-line systems compared to that of strong systems.

Figure~\ref{fig:figabun} also shows average relative abundances from absorbers such as DLAs, Mg II absorbers and 2DAs. The 2DA element abundances are generally more heavily depleted than either of the Ca II absorbers while the Mg II absorber and DLA's element relative abundances are generally less depleted than either of the Ca II absorbers. These dust depletion patterns indicate that  Ca II absorbers generally have environments with higher dust contents than DLAs and Mg II absorbers, but with lower dust contents than 2DAs. 

For both subpopulations in our catalog, Si (an $\alpha$ element synthesized by Type II supernovae) is noticeably more depleted than the Milky Way's components and also Sardane et al's sample. This may be due to the higher redshift nature of our catalog than Sardane's catalog.  Our catalog, when compared to \citet{2014MNRAS.444.1747S}'s catalog, has a much larger proportion of high redshift Ca II absorbers, as seen in Figure \ref{fig:figzabs}, and a greater median redshift of $\overline{z_{\text{abs}}}=0.85$ compared to \citet{2014MNRAS.444.1747S}'s catalog, which has a median redshift of $\overline{z_{\text{abs}}}=0.59$.

\subsection{2175 \r{A} Dust Absorbers (2DAs)} \label{sssec:2DA}
\begin{figure}
    \centering
    \includegraphics[width =\linewidth]{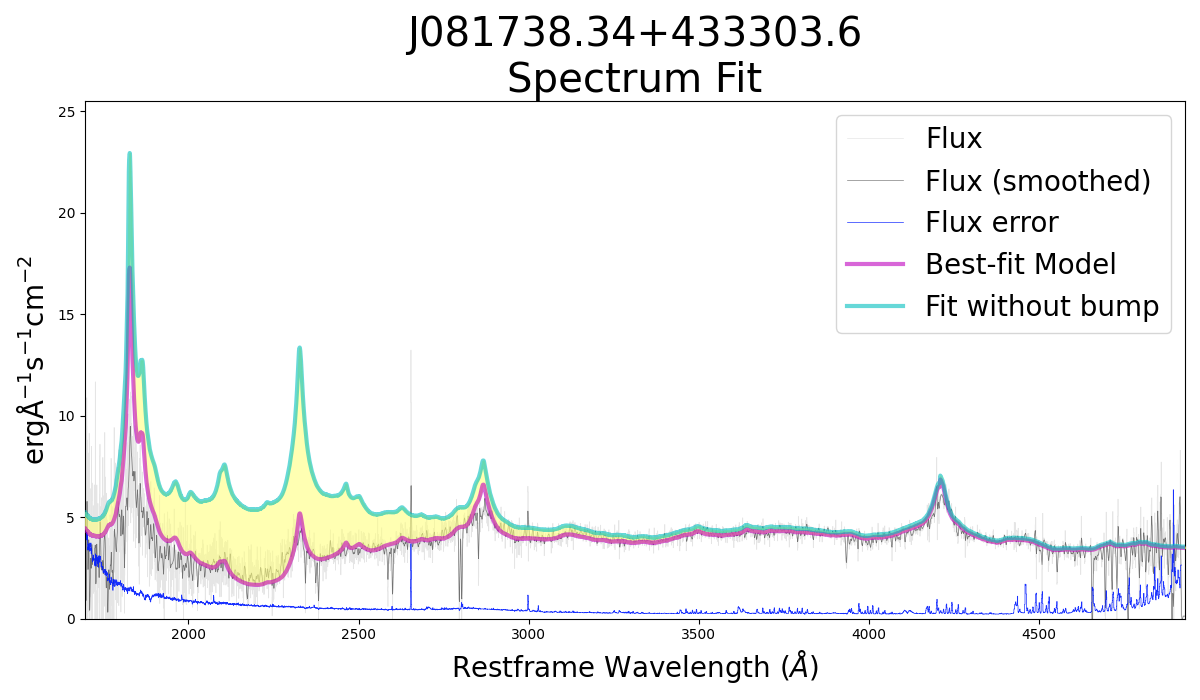}
    \includegraphics[width =\linewidth]{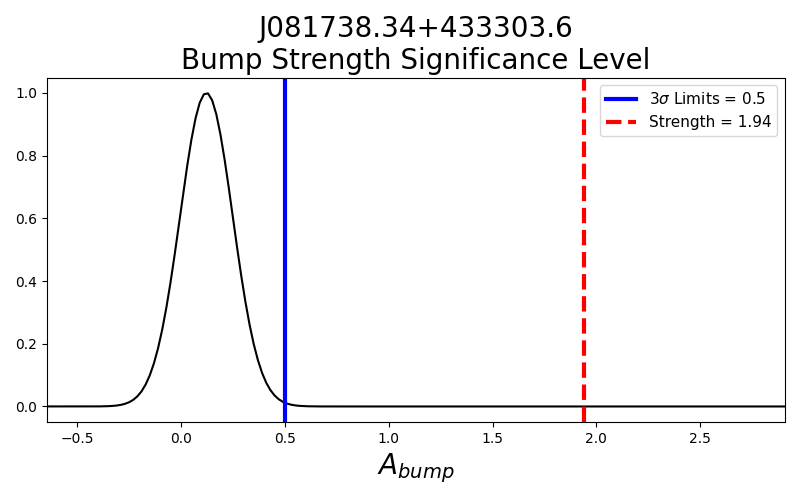}
    \caption{Top: Extinction curve fit for J081738.34+433303.6. The magenta fit is the best fitting model. The cyan model is only the linear component of the extinction, fitted without the bump factored in. Bottom: Bump strength significance plot for J081738.34+433303.6. As the strength of the 2175\AA\ dust bump is greater than the $3\sigma$ limits, this candidate is a discovered 2DA.}
    \label{fig:DAfit}
\end{figure}

\begin{figure}
    \centering
    \includegraphics[width =\linewidth]{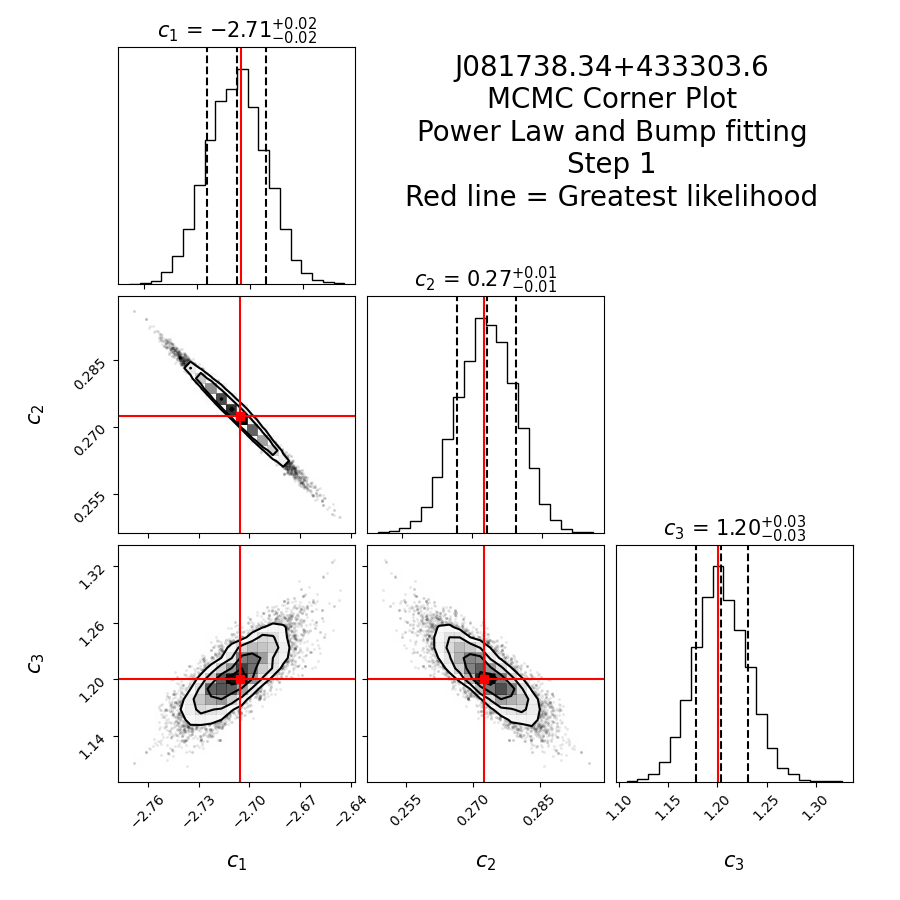}
    \includegraphics[width =\linewidth]{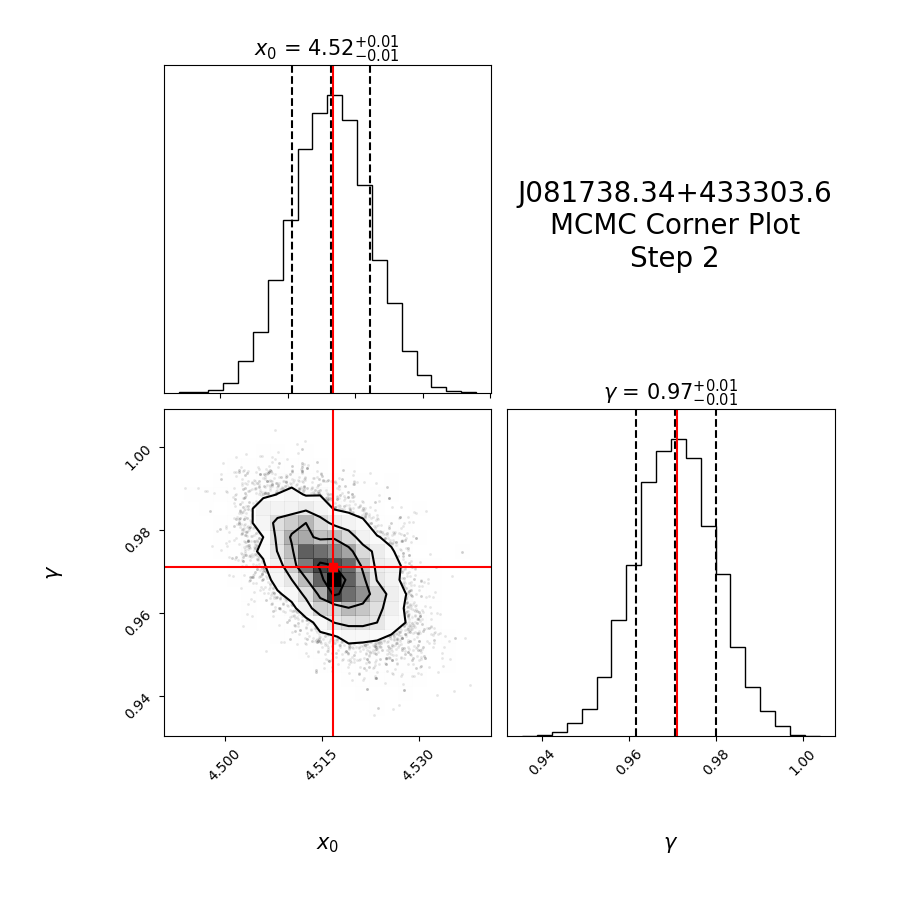}
    \caption{Corner plot showing the normal posterior probability distributions from our MCMC parameter solver for a 2DA extinction fitting of the quasar spectrum, J081738.34+433303.6.}
    \label{fig:cornerplot}
\end{figure}

We discovered 12 2DAs in our catalog. Our Ca II absorber catalog contains 100 absorption-line systems with $z_{abs} \geq 0.8$, all possible  2DA candidates based on redshift. Extinction curves with Drude profiles were solved and fitted for the 2DA candidates’ spectra following the procedures described in Section 2. Figure \ref{fig:DAfit} shows an example extinction curve fitting and a plot evaluating the area of the Drude profile (bump strength). An example corner plot showing the best fit parameters for a 2DA from our MCMC solver is shown in Figure \ref{fig:cornerplot}. Candidates were disregarded if the bump significance level was less than $3\sigma$. Candidates with the bump significance level $\geq3\sigma$ are considered as dust 2175 \AA\ absorbers or 2DAs. The best fitting parameters along with estimated visible extinction A\textsubscript{V} values (magnitudes) for our 2DAs' extinction curves can be found in Table \ref{tab:Table 3}. Note that the low redshift of our 2DAs puts the V-band along the dust bump feature, which is why the reported A\textsubscript{V} values are so large.

\begin{table*}
\centering
\caption{\label{tab:Table 3} Best-fit parameters of the extinction curves for our 2DAs. Reported parameters for each 2DA are averaged between the MW and LMC derived values (see section \ref{sssec:2DASearch}). Reported errors incorporate statistical uncertainties from the MCMC fittings and difference in MW and LMC derived values.}
\begin{tabular}[t]{ccccccccc} 
\toprule
{Quasar} & {$z_{\text{abs}}$} & {$c_1$} & {$c_2$} & {$c_3$} & {$x_0$} & {$\gamma$} & {A\textsubscript{bump}} & {$A_V$}\\
\midrule
{J005251.81+090945.2}&{1.32}&{$-3.96\pm0.03$}&{$0.34\pm0.01$}&{$1.07\pm0.06$}&{$4.48\pm0.01$}&{$0.94\pm0.01$}&{$1.79\pm0.18$}&{1.52}{$\pm0.20$}\\
{J081738.34+433303.6}&{1.10}&{$-2.7\pm0.01$}&{$0.26\pm0.01$}&{$1.25\pm0.06$}&{$4.52\pm0.01$}&{$0.99\pm0.01$}&{$1.98\pm0.12$}&{1.21$\pm0.19$}\\
{J120301.01+063441.5}&{0.86}&{$-2.8\pm0.01$}&{$0.31\pm0.01$}&{$0.69\pm0.02$}&{$4.52\pm0.04$}&{$0.97\pm0.01$}&{$1.11\pm0.28$}&{1.54}{$\pm0.15$}\\
{J121219.87+293622.8}&{1.22}&{$-3.84\pm0.01$}&{$0.07\pm0.00$}&{$0.28\pm0.01$}&{$4.70\pm0.01$}&{$0.85\pm0.02$}&{$0.52\pm0.11$}&{3.18}{$\pm0.31$}\\
{J141850.71+154106.3}&{1.04}&{$-3.79\pm0.01$}&{$0.07\pm0.00$}&{$0.45\pm0.02$}&{$4.90\pm0.01$}&{$0.86\pm0.02$}&{$0.82\pm0.12$}&{3.20}{$\pm0.31$}\\
{J141951.84+470901.3}&{1.27}&{$-4.47\pm0.02$}&{$0.19\pm0.01$}&{$0.19\pm0.01$}&{$4.32\pm0.01$}&{$0.92\pm0.04$}&{$0.32\pm0.09$}&{3.22}{$\pm0.51$}\\
{J154457.53+283325.6}&{1.07}&{$-1.00\pm0.01$}&{$-0.02\pm0.00$}&{$0.36\pm0.02$}&{$4.74\pm0.02$}&{$0.75\pm0.01$}&{$0.74\pm0.10$}&{0.94}{$\pm0.25$}\\
{J110236.85+235706.5}&{1.36}&{$-2.96\pm0.01$}&{$0.21\pm0.01$}&{$0.30\pm0.01$}&{$4.69\pm0.01$}&{$0.90\pm0.01$}&{$0.52\pm0.08$}&{1.72}{$\pm0.13$}\\
{J135123.47+474712.1}&{1.06}&{$-3.32\pm0.01$}&{$0.04\pm0.00$}&{$0.29\pm0.01$}&{$4.74\pm0.01$}&{$0.82\pm0.01$}&{$0.55\pm0.010$}&{2.88}{$\pm0.11$}\\
{J082516.44+094700.4}&{1.29}&{$-2.01\pm0.0$}&{$-0.14\pm0.00$}&{$0.34\pm0.01$}&{$4.71\pm0.01$}&{$0.86\pm0.01$}&{$0.61\pm0.07$}&{2.24}{$\pm0.08$}\\
{J082516.44+094700.4}&{1.11}&{$-3.58\pm0.0$}&{$-0.02\pm0.00$}&{$0.26\pm0.01$}&{$4.88\pm0.01$}&{$0.67\pm0.01$}&{$0.61\pm0.09$}&{3.33}{$\pm0.30$}\\
{J124914.78+301550.0}&{1.27}&{$-2.55\pm0.01$}&{$0.34\pm0.02$}&{$0.50\pm0.02$}&{$4.68\pm0.01$}&{$0.92\pm0.01$}&{$0.86\pm0.09$}&{0.87}{$\pm0.08$}\\
\midrule
{Average}&{-}&{$-3.08\pm-0.04$}&{$0.14\pm0.03$}&{$0.50\pm0.05$}&{$4.66\pm0.05$}&{$0.87\pm0.05$}&{$0.87\pm0.51$}&{2.34}{$\pm0.22$}\\
\bottomrule
\end{tabular}
\end{table*}

\begin{figure}
    \centering
    \includegraphics[width =\linewidth]{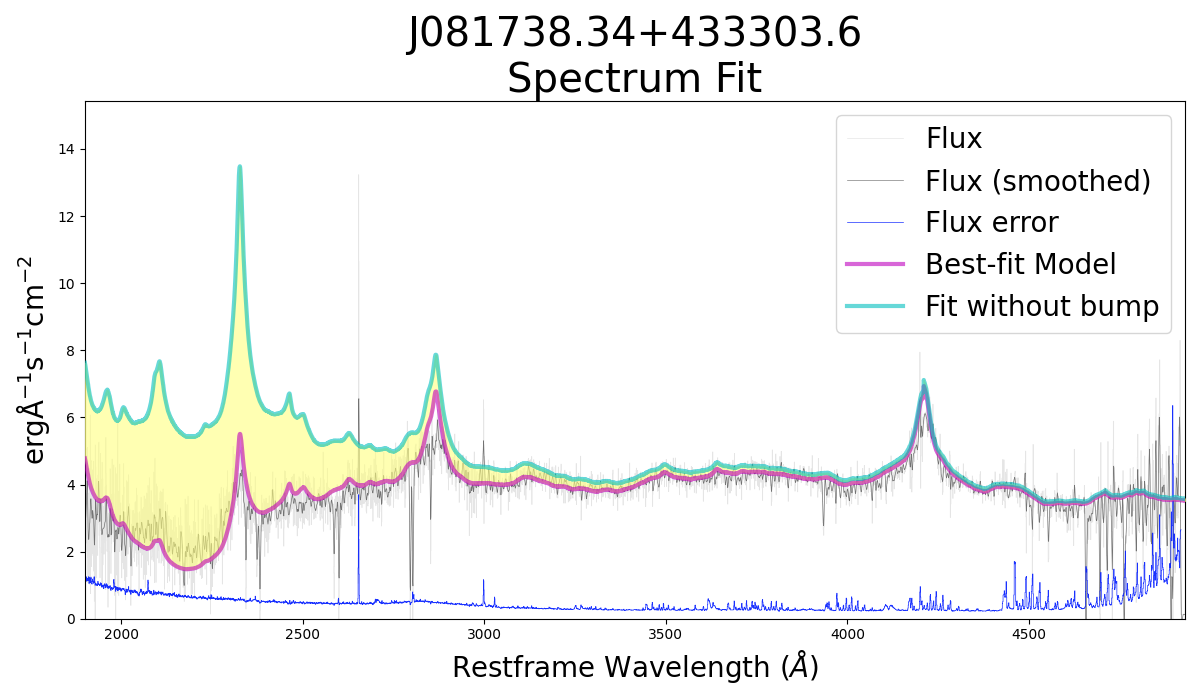}
    \includegraphics[width =\linewidth]{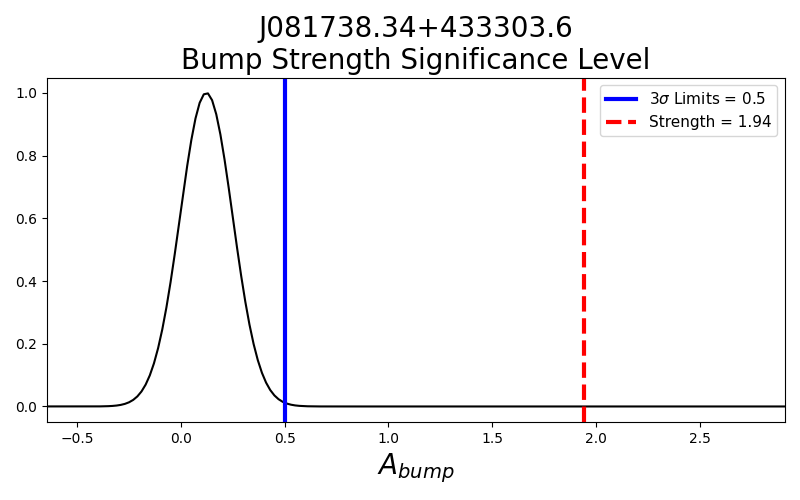}
    \caption{Top: Extinction curve fit for J081738.34+433303.6, the same dust absorber as Figure \ref{fig:DAfit}, but performed with the spectrum cut at $\lambda$\textsubscript{restframe} = 1900 \AA. The magenta fit is the best fitting model. The cyan model is only the linear component of the extinction, fitted without the bump factored in. Bottom: Bump strength significance plot for J081738.34+433303.6. The calculated area of the 2175 \AA\ dust bump is 1.94, the same as that from the uncut spectrum. Both bump strengths are much greater than the $3\sigma$ limits of 0.5.}
    \label{fig:figcutspec}
\end{figure}

The SDSS quasar spectra of absorption-line systems feature different wavelength range coverage in the optical spectral window covered by the SDSS original spectrograph (for DR7 data) or SDSS-III BOSS spectrograph (for DR12 data) due to different absorber's redshifts. Quasar spectra with 2DAs at lower redshift regime (such as $z_{abs}\sim 0.8-1.0$) may not completely cover the broad 2175 \AA\ dust bump's features, which may affect our bump fitting accuracy. We investigated the effectiveness of our FM parameterized extinction curve fitting program to quasar spectra with lower redshift 2DAs by comparing the fitting results of quasar spectra with a full wavelength coverage of the 2175 \AA\ bump at higher redshifts with the same spectra cut off at the absorber's rest frame 1900 \r{A} to mimic absorbers at lower redshifts. An example of this is shown in Figure~\ref{fig:figcutspec}, displaying the results of J081738.34+433303.6's spectrum when cut in comparison to its uncut fitting from Figure~\ref{fig:DAfit}. Our study results show little variation in bump measurement strengths between smaller and larger wavelength coverage. This demonstrates the validity of our extinction fitting program to measure 2175 \AA\ dust bump strengths for all quasar spectra with 2DAs  in the $1.4\geq z\geq 0.8$ range in our sample.

\begin{figure}
    \centering
    \includegraphics[width =\linewidth]{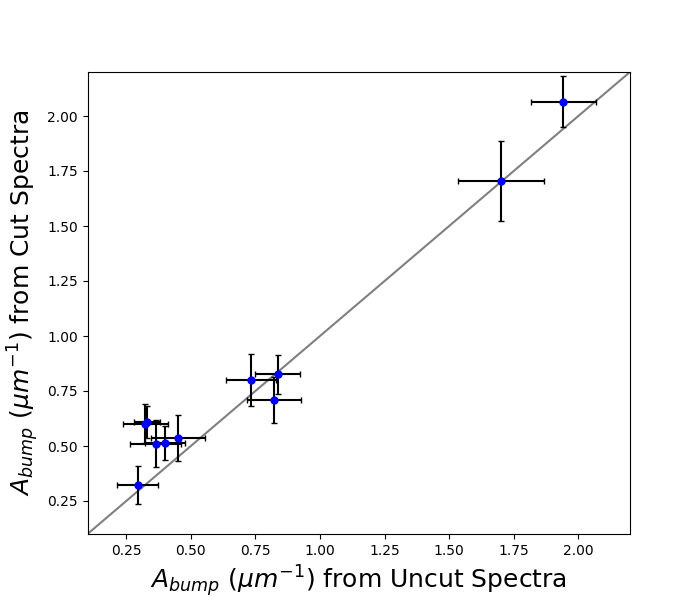}
    \caption{A comparison of calculated area of the 2175\AA\ dust bump (strength) from uncut spectra vs spectra cut at the rest frame of 1900 \AA\ for 11 2DAs with $z_{\text{abs}} > 1.0)$ in our sample. The black line represents a 1:1 comparison. There is an average $10\%$ difference in bump strength between the cut and uncut spectra.}
    \label{fig:abump}
\end{figure}

We evaluated 11 of our 12 dust absorbers, all of which with $z>1$. Evaluations showed little variation in bump strength between fitting of the uncut and cut spectra with an average 10\% difference in bump strength (area of drude profile). Figure \ref{fig:abump} exhibits the bump strength difference between the cut and uncut spectra for our 2DAs at $z_{abs}>1.0$. This demonstrates that the inclusion of a range of lower wavelengths has no significant effect on our bump strength measurement accuracy. Thus, our fitting program can fairly accurately measures the bump strength of the 2175 \r{A} dust feature at $z_{abs}\sim 0.8-1.0$ where the broad 2175 \AA\ bump features are not completely covered by the SDSS quasar spectra.

The sample of 12 2DAs is composed of 7 strong Ca II absorption-line systems and 5 weak Ca II absorption-line systems. The strong population as a whole appears to be dustier than the weak population. Of the redshift eligible systems, 33\% of strong Ca II absorption-line systems shows the 2175\AA\ dust bump, while only 6.3\% of the eligible weak Ca II absorption-line systems shows the bump. Such stark population distinctions in terms of the 2175 \AA\ dust absorption feature are consistent with our results regarding distinctions in element depletion and environment. 

\begin{figure*}
    %\centering
    \includegraphics[width =\linewidth]{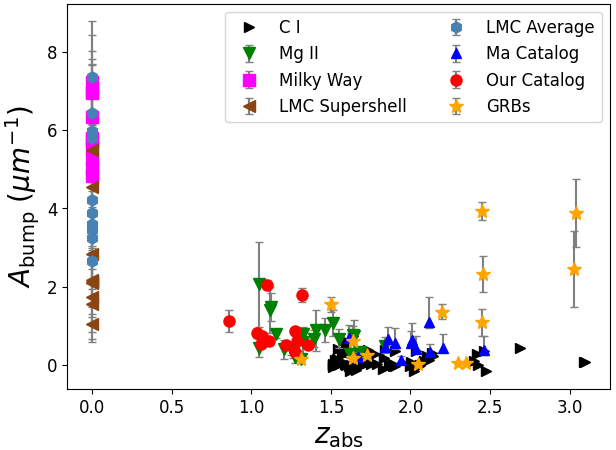}
    \caption{Absorber redshift versus 2175 \AA\ bump strength for dust absorber populations. The black triangles are 2175 \AA\ dust bumps found in C I absorbers \citep{2015A&A...580A...8L} and the green triangles are dust bumps found in Mg II absorbers \citep{2011ApJ...732..110J}. The magenta squares are MW dust bumps \citep{2007ApJ...663..320F}, the brown triangles are the bumps in the LMC supershell, and the steel blue hexagons are the average LMC bumps \citep{2003ApJ...594..279G}. The blue triangles are 2175 \AA\ bumps from the \citet{2018MNRAS.474.4870M} catalog and the red circles are dust bumps found in Ca II absorbers from our catalog. The yellow stars are 2175 \AA\ bumps from GRB optical afterglow spectra \citep{2011A&A...532A.143Z, 2009ApJ...697.1725E, 2019MNRAS.486.2063H}. The plot exhibits a general decrease in 2175 \AA\ bump strength as absorber redshift increases. This negative correlation is strong with $R=-0.74$. This indicates redshift evolution of dust, implying dust growth with time in the universe.}
    \label{fig:da_zabs}
\end{figure*}

We also compiled 142 2DAs from literature to compare absorber redshift and 2175 \AA\ dust bump strength. With 2175 \AA\ dust bumps from C I absorbers \citep{2015A&A...580A...8L}, Mg II absorbers \citep{2011ApJ...732..110J}, Ca II absorbers (our catalog), the Milky Way \citep{2007ApJ...663..320F}, the average LMC and LMC supershell \citep{2003ApJ...594..279G}, and \citet{2018MNRAS.474.4870M}'s catalog, we plotted absorber redshift versus 2175 \AA\ bump strength in Figure \ref{fig:da_zabs}. Additionally, we compiled 14 2DAs detected from the optical afterglow spectra of gamma-ray bursts (GRBs) \citep{2011A&A...532A.143Z,2009ApJ...697.1725E, 2019MNRAS.486.2063H}. We will refer to these as GRB 2DAs and the 2DAs detected from quasar spectra as quasar 2DAs.

We note that Figure \ref{fig:da_zabs} features a gap between the local group and z=0.83 because there have been only three solid detections of the 2175 \AA\ dust bump in individual extragalactic environments prior to the present study \citep{2002ApJ...574..719M, 2004ApJ...614..658J, 2006MNRAS.372L..38E}. None of those studies reported bump strength nor utilized FM parameterization to fit their extinction curves. Without the FM parameters, we cannot derive the area of the 2175 \AA\ dust bump (bump strength, A\textsubscript{bump}) and cannot include those measurements into the plot.

Figure \ref{fig:da_zabs} exhibits a general decrease in 2175 \AA\ bump strength as absorber redshift increases. Correlational analysis shows that there is a strong negative correlation between absorber redshift and local group 2DA and quasar 2DA dust bump strength with correlation coefficient $R=-0.82$. This implies that the 2175 \AA\ dust bump tends to be stronger when it is closer to us, indicating dust bump evolution with redshift and implying global dust and metal enrichment history. For the first time, we find indication of dust growth with time in the universe. In order to confirm this global trend, we have added 2DA measurements from totally independent followup observations of high-z Gamma Ray Burst optical afterglows in this figure. A majority of the added GRB 2DAs also follow this trend except for a few high redshift GRB 2DAs. This may indicate that those GRB 2DAs possibly trace highly dusty regions of galaxies hosting GRBs. Further followup observations of these sources may reveal their enrichments. Regardless, high-z GRB 2DAs appear to show a relatively larger range of UV bump strengths than quasar 2DAs. Correlational analysis of the global dust trend, taking local group 2DAs, quasar 2DAs, and GRB 2DAs into account yielded a strong negative correlation of $R=-0.74$ between absorber redshift and 2DA bump strength.

Future investigations of the chemical properties of the Ca II absorbers containing this 2175 \r{A} broad absorption feature in their quasar spectra can potentially provide unique insights into the galactic evolution that led to today’s Milky Way which has strong 2175 \AA\ dust absorption bump. Since dust producing these 2175 \AA\ dust absorption bump  likely consists of organic PAH molecules (e.g., \citealt{2003ARA&A..41..241D}), which can be potentially used to trace assembly of organic molecules in the early universe and  possibly the beginning of life itself \citep{2004ApJ...609..589W, 2018MNRAS.474.4870M}. 

\subsection{Ca II Absorber Reddening}\label{sssec:reddening}
To measure the average reddening of Ca II absorbing gas, we compared unnormalized flux composites of Ca II absorbers to unabsorbered references. To create unabsorbed reference composites, we found a quasar spectrum without absorbers for each Ca II absorber in our catalog. The sample of these spectra will be referred to as the non-absorber sample. Identification of non-absorber matches required nearly identical SDSS \emph{i}-band magnitudes and emission redshifts (z\textsubscript{em}). Nearly indistinguishable emission redshifts also allow us to use the known absorber redshifts of our Ca II absorbers to create a nearly identical composite spectrum without reddening produced by Ca II absorbers when shifting non-absorber matches to the rest frame of absorbers.

Search for non-absorber matches began by finding quasar spectra with $\abs{\Delta i}\leq0.2$, $\abs{\Delta z_{\text{em}}}\leq0.01$, no intervening Mg II absorption (cross referenced Mg II catalogs for SDSS DR 7 and 12 from \citep{2013ApJ...770..130Z, 2019MNRAS.487..801Z}, and no known broad absorption lines (using published SDSS catalogs) for each Ca II absorber. Only the SDSS data releases from which the referenced Ca II absorber was found was searched. Spectra meeting these requirements were reviewed for absorption transitions for a homogeneous non-absorber sample. If there was more than one quasar spectrum that fulfilled both the requirements and visual inspection, the non-absorber match was decided by the spectrum with the smallest $(\Delta z_{\text{em}}/R_{z_{\text{em}}})^2+(\Delta i/R_i)^2$ result ($R$ is the rms). This approach follows ones taken in \citet{2015MNRAS.452.3192S} and \citet{2006MNRAS.367..945Y}. 

Unnormalized flux composites were then created for the $W_0^{\lambda3934}<0.7$  \AA, $W_0^{\lambda3934}\geq0.7$ \AA, and 2DA Ca II absorber subsamples. Then non-absorber compsites were created for each aforementioned sample with their respective non-absorber matches. Composites were formed by taking the geometric mean of unnormalized flux spectra shifted to restframe as we followed Sardane et al.'s (2015) approach to measuring reddening. Continua were fit for each composite using a fifth-degree polynomial. The parameter $\mathcal{R}$ was calculated by the absorbed continuum to non-absorbed continuum ratio at the rest wavelength of 2200 \r{A}. To calculate error of each composite's continuum value at 2200\AA\, we measured the rms of each composite within a 100 \AA\ window of 2200 \AA. Then we propagated the error of the continua to calculate the error of $\mathcal{R}$.

\begin{figure}
    \centering
    \includegraphics[width =\columnwidth]{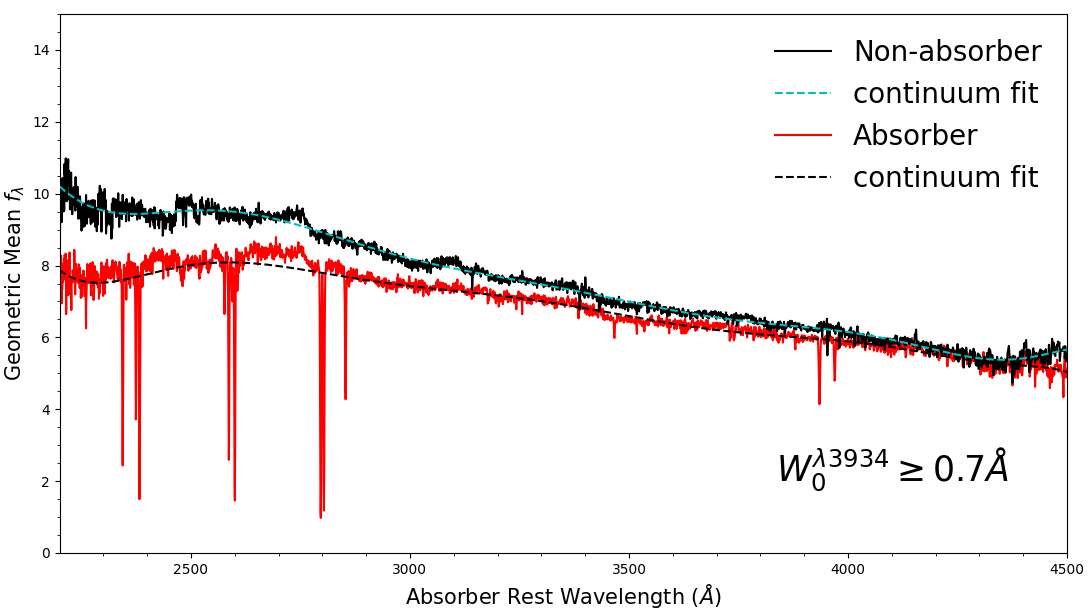}
    \includegraphics[width =\columnwidth]{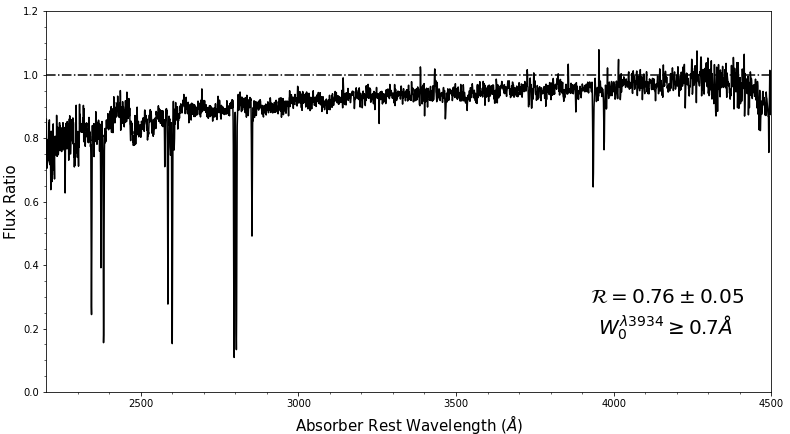}
    \caption{Top: Ca II absorber (red) and matched non-absorber (black) geometric mean  restframe composites for the $W_0^{\lambda3934}\geq0.7$ \AA\ strong subpopulation. Prominent Ca II, Fe II, Mg II and Mg I absorption lines are visible in the absorber composite. Bottom: The ratio of absorber composite flux to non-absorber composite flux. The calculated ratio $\mathcal{R}$ at 2200 \AA\ is $0.76\pm0.05$.}
    \label{fig:figreddS}
\end{figure}

\begin{figure}
    \centering
    \includegraphics[width =\columnwidth]{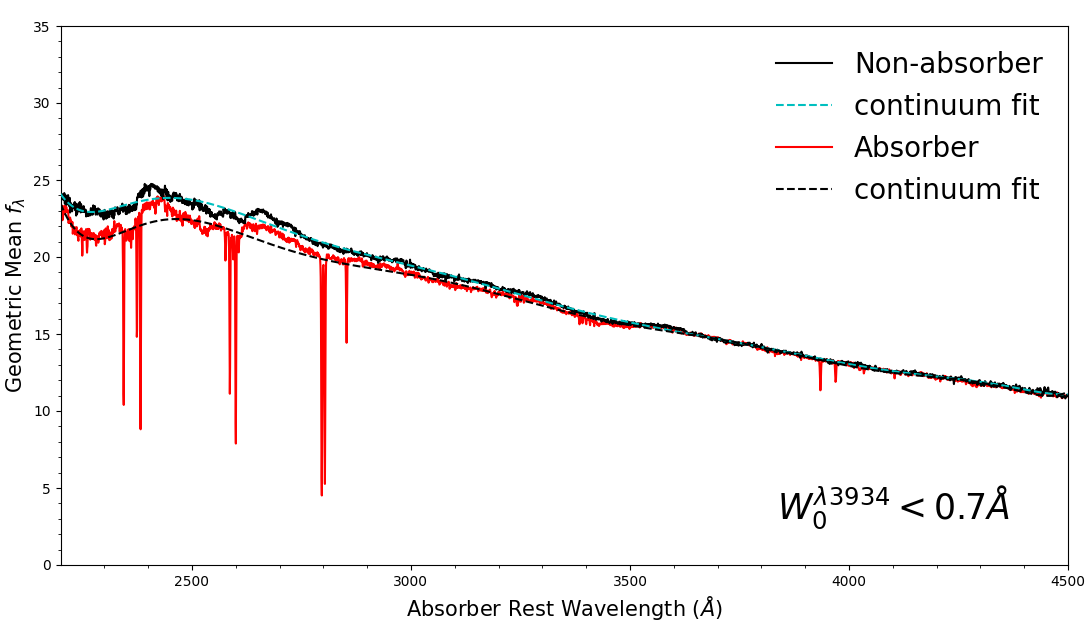}
    \includegraphics[width =\columnwidth]{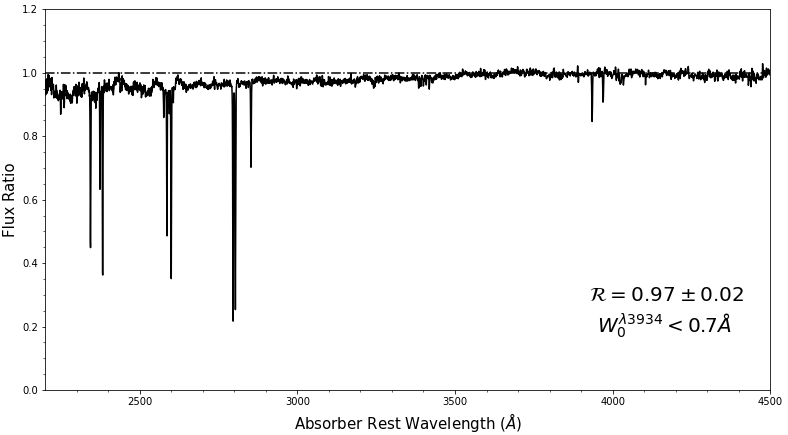}
    \caption{Top: Ca II absorber (red) and matched non-absorber (black) geometric mean  restframe composites for the $W_0^{\lambda3934}<0.7$ \AA\ weak subpopulation. Prominent Ca II, Fe II, Mg II and Mg I absorption lines are visible in the absorber composite. Bottom: The ratio of absorber composite flux to non-absorber composite flux. The calculated ratio $\mathcal{R}$ at 2200 \AA\ is $0.97\pm0.02$, which is greater than that of the strong subpopulation's  $\mathcal{R}=0.76\pm0.05$.}
    \label{fig:figreddW}
\end{figure}

\begin{figure}
    \centering
    \includegraphics[width =\columnwidth]{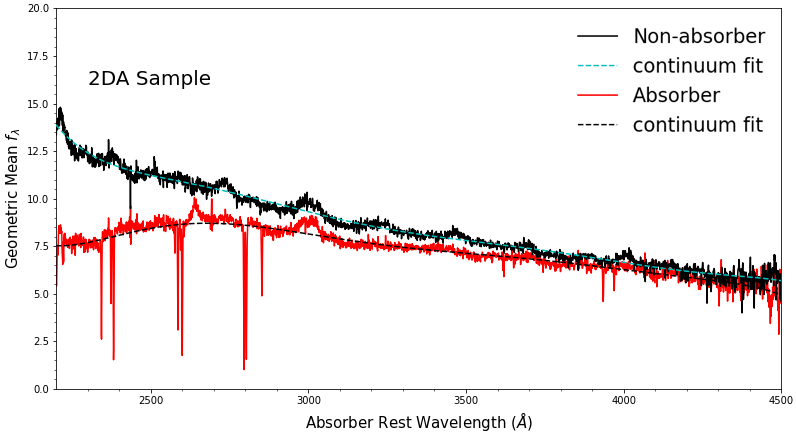}
    \includegraphics[width =\columnwidth]{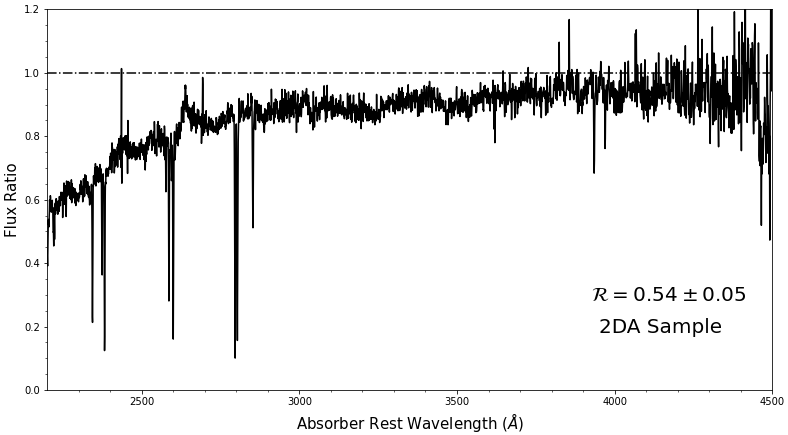}
    \caption{Top: Ca II absorber (red) and matched non-absorber (black) geometric mean  restframe composites for the dust absorber subpopulation. Prominent Ca II, Fe II, Mg II and Mg I absorption lines are visible in the absorber composite. Bottom: The ratio of absorber composite flux to non-absorber composite flux. The calculated ratio $\mathcal{R}$ at 2200 \AA\ is $0.54\pm0.05$, which is less than that of both the strong subpopulation's $\mathcal{R}=0.76\pm0.05$ and the weak population's $\mathcal{R}=0.97\pm0.02$. The absorption and emission features still present in the composites, due to the small number of 2DA spectra combined, cause the emission looking features in the flux ratio.}
    \label{fig:figredd_da}
\end{figure}

The matched composites and their continua for the subsamples are shown in Figure~\ref{fig:figreddS} for the strong subpopulation, Figure~\ref{fig:figreddW} for the weak subpopulation, and Figure~\ref{fig:figredd_da} for the 2DA sample. The figure also shows the flux ratios for each subsample. Note that the dust absorber composite is only composed of 12 spectra which is why there are still certain absorption or emission features present. Both the strong and weak absorber composites are free of such features because of their much larger sample size.


The strong Ca II absorber subpopulation is more heavily reddened than the weak subpopulation with an $\mathcal{R}$ value of $\mathcal{R}=0.76\pm0.05$ (Figure \ref{fig:figreddS}) compared to the weak subpopulation's $\mathcal{R}=0.97\pm0.02$ (Figure \ref{fig:figreddW}). This result is expected because the strong subpopulation demonstrates a heavier dust depletion pattern than the weak subpopulation (Section 3.4). The dust absorber subpopulation shows the smallest $\mathcal{R}$ value with $\mathcal{R} = 0.54\pm0.05$, indicating the most dust reddening among these three samples as expected (Figure \ref{fig:figredd_da}).

\begin{table}
\centering
\caption{\label{tab:R} Reddening results for our three Ca II absorber subsamples. Results from our catalog are presented alongside results from \citet{2015MNRAS.452.3192S}. The parameter $\mathcal{R}$ is the absorbed to unabsorbed flux ratio at the rest frame of 2200 \AA. Our results show that our catalog is slightly less reddened than \citet{2015MNRAS.452.3192S}'s, but as our results are within $1\sigma$ of theirs, there is no significant difference in reddening.}
\begin{tabular}[t]{ccc} 
\toprule
\multicolumn{3}{|c|}{$\mathcal{R}$} \\
{Subsample} & {Our Catalog} & {Sardane et al. (2015)}\\
\midrule
{$W_0^{\lambda3934}<0.7$\AA}&{$0.97\pm0.02$}&{0.95}\\
{$W_0^{\lambda3934}\geq0.7$\AA}&{$0.76\pm0.05$}&{0.73}\\
{Dust Absorbers}&{$0.54\pm0.05$}&{-}\\
\bottomrule
\end{tabular}
\end{table}

Compared to \citet{2015MNRAS.452.3192S} reported reddening for their 2014 catalog, our catalog has slightly higher $\mathcal{R}$ value of $0.97\pm0.02$ for the $W_0^{\lambda3934}<0.7$ \AA\ subsample compared to \citet{2015MNRAS.452.3192S}'s $\mathcal{R}=0.95$, but our results are ultimately comparable because they are within $1\sigma$ of \citet{2015MNRAS.452.3192S}'s. Our strong sample with the $\mathcal{R}$ value of $0.76\pm0.05$ is also comparable to \citet{2015MNRAS.452.3192S}'s 0.73. These comparisons are also reported in Table \ref{tab:R}. While \citet{2015MNRAS.452.3192S} did not search for dust absorbers within their catalog, our dust absorbers' $\mathcal{R}$ value is significantly lower than any of their reported $\mathcal{R}$ values and our values for both Ca II absorber samples, indicating more dust in these 2DAs. Our catalog slighly less reddening for both Ca II absorber samples may be attributed a higher redshift distribution as the majority of our catalog (61\%) are high redshift absorbers while only 23\% of \citet{2015MNRAS.452.3192S}'s catalog are such. Due to global chemical evolution, we expect that Ca II absorbers at lower redshifts have more dust produced by stars and interstellar processes that those at higher redshifts. However, as our catalog's results are within $1\sigma$ of \citet{2015MNRAS.452.3192S}'s, the difference in reddening may be attributed to small flux fluctuations of different samples.

\section{Discussion}
From our comparison of abundance ratios of several elements (Si, Mn, Cr, Fe, Ni, Ti, Ca) to Zn  in our Ca II absorber's sample to that of \citet{2014MNRAS.444.1747S}'s sample and various Milky Way interstellar medium components, our sample's heavy depletion of Si was a stand out result. This may be caused by different nucleosynthesis history and/or chemical evolution  of these Ca II absorbers.  More data from Ca II absorbers and other absorbers should be collected to study this possible evolution trend. 

Our curve of growth analysis of each Ca II absorber subpopulation resulted in the $W_0^{\lambda3934}\geq0.7$ strong population having a larger b value Doppler parameter of 63.61 km/sec than the $W_0^{\lambda3934}<0.7$ weak population's b value Doppler parameter of 46.29 km/sec. This implies more velocity components in the strong population's environment than the weak population's. 

\citet{2015MNRAS.452.3192S}’s studies could not distinguish between the accuracy of fitting the LMC and SMC dust law to the extinction of quasar Ca II absorbers from both populations. However, our discovery of significant 2175 \r{A} dust bumps in certain Ca II absorption-line systems suggest that some of the Ca II absorbers, especially the strong population, may have substantial amount of PAH type dust in their gaseous environment responsible for the LMC type 2175 \AA\ dust absorption bump. 

Interstellar dust is vital to understanding galaxy formation and evolution. However, it is often studied at high redshifts, and investigation of the more recent universe is necessary for understanding evolution. The unique ability of Ca II absorbers to probe low redshifts allows for the analysis of dust in galaxies and around galaxies in the more recent Universe. Revealing both the chemical abundance and dust content of low redshift intervening systems, Ca II absorbers provide a robust understanding of the chemical enrichment of ISMs, CGMs, and IGMs. Ca II absorbers are also an important class of quasar absorbers because of the variety of H I column densities they exhibit. With the addition of our new discoveries, there are now over 650 known Ca II absorbers, a sample size large enough to accurately measure global star formation rates with H I column densities at low redshifts.

While use of Mg II absorbers to discover Ca II absorbers is more efficient than an attempt at manual detection without such aids, an automated discovery process using a neural network \citep{2022MNRAS} appears to be able to detect more Ca II absorbers in DR7 and DR12 than our manual method. We plan to analyze more Ca II absorbers from our deep neural network and also Sardane et al's sample to help improve statistics results drawn from this study. 

\section{Conclusion}
We detected a total of  165 new Ca II quasar absorbers in the SDSS DR7 and DR12 data. Although our catalog size is the second largest one compared to Sardane's, our catalog is unique from other large Ca II catalogs due to its large proportion of absorbers in the high redshift regime.  The previous discovery of two distinct strong and weak populations in quasar Ca II absorbers was confirmed by our independent measurement results, indicating different environments associated with each population. Strong Ca II absorbers show a large dust depletion pattern consistent with that of gaseous components from the disc-like and halo-like environment in the Milky Way while weak Ca II absorbers show a small dust depletion pattern consistent with that of gaseous components in the halo environment in the Milky Way. In addition, strong absorbers show stronger dust reddening than weak Ca II absorbers. Comparing dust depletion patterns of Ca II absorbers with that of DLAs, Mg II absorbers,  and 2DAs show that  Ca II absorbers generally have environments with higher dust contents than DLAs and Mg II absorbers, but with lower dust contents than 2DAs. 

Our search for 2175 \r{A} dust absorbers using Ca II absorbers, the study to do so, resulted in the detection of 12 2DAs. About 33\% of strong Ca II absorbers show the 2175 \AA\ absorption bump while only about 6\% of weak Ca II absorbers show this kind of dust absorption bump.  Analysis results of these 2DAs show their spectra are much reddened than other Ca II absorbers without 2DAs. 

Compiling our 2DA sample with 142 other 2DAs from literature, we compared absorber redshift to 2175 \AA\ dust bump strength, finding a strong negative correlation. Figure \ref{fig:da_zabs} exhibits the implied overall dust evolution trend. This significant discovery indicates dust growth in the universe over time. This finding also supports the global chemical enrichment of galaxies with more dust accumulated in lower redshift galaxies.

\section{Acknowledgements}
Funding for SDSS-III has been provided by the Alfred P. Sloan Foundation, the Participating Institutions, the National Science Foundation, and the U.S. Department of Energy Office of Science. The SDSS-III web site is http://www.sdss3.org/.

SDSS-III is managed by the Astrophysical Research Consortium for the Participating Institutions of the SDSS-III Collaboration including the University of Arizona, the Brazilian Participation Group, Brookhaven National Laboratory, Carnegie Mellon University, University of Florida, the French Participation Group, the German Participation Group, Harvard University, the Instituto de Astrofisica de Canarias, the Michigan State/Notre Dame/JINA Participation Group, Johns Hopkins University, Lawrence Berkeley National Laboratory, Max Planck Institute for Astrophysics, Max Planck Institute for Extraterrestrial Physics, New Mexico State University, New York University, Ohio State University, Pennsylvania State University, University of Portsmouth, Princeton University, the Spanish Participation Group, University of Tokyo, University of Utah, Vanderbilt University, University of Virginia, University of Washington, and Yale University.

The authors would like to thank the anonymous referee who has provided valuable suggestions which helped improve the quality of this paper.

\section{Data Availability Statement}
The data underlying this article was accessed from the Sloan Digital Sky Survey (https://classic.sdss.org). The data generated from this research are available in the article and in its online supplementary material.


\bibliographystyle{mnras}
\bibliography{mnras_ref}

\begin{thebibliography}{}
\makeatletter
\relax
\def\mn@urlcharsother{\let\do\@makeother \do\$\do\&\do\#\do\^\do\_\do\%\do\~}
\def\mn@doi{\begingroup\mn@urlcharsother \@ifnextchar [ {\mn@doi@}
  {\mn@doi@[]}}
\def\mn@doi@[#1]#2{\def\@tempa{#1}\ifx\@tempa\@empty \href
  {http://dx.doi.org/#2} {doi:#2}\else \href {http://dx.doi.org/#2} {#1}\fi
  \endgroup}
\def\mn@eprint#1#2{\mn@eprint@#1:#2::\@nil}
\def\mn@eprint@arXiv#1{\href {http://arxiv.org/abs/#1} {{\tt arXiv:#1}}}
\def\mn@eprint@dblp#1{\href {http://dblp.uni-trier.de/rec/bibtex/#1.xml}
  {dblp:#1}}
\def\mn@eprint@#1:#2:#3:#4\@nil{\def\@tempa {#1}\def\@tempb {#2}\def\@tempc
  {#3}\ifx \@tempc \@empty \let \@tempc \@tempb \let \@tempb \@tempa \fi \ifx
  \@tempb \@empty \def\@tempb {arXiv}\fi \@ifundefined
  {mn@eprint@\@tempb}{\@tempb:\@tempc}{\expandafter \expandafter \csname
  mn@eprint@\@tempb\endcsname \expandafter{\@tempc}}}

\bibitem[\protect\citeauthoryear{{Abazajian} et~al.,}{{Abazajian}
  et~al.}{2009}]{2009ApJS..182..543A}
{Abazajian} K.~N.,  et~al., 2009, \mn@doi [\apjs]
  {10.1088/0067-0049/182/2/543}, \href
  {https://ui.adsabs.harvard.edu/abs/2009ApJS..182..543A} {182, 543}

\bibitem[\protect\citeauthoryear{{Alam} et~al.,}{{Alam}
  et~al.}{2015}]{2015ApJS..219...12A}
{Alam} S.,  et~al., 2015, \mn@doi [\apjs] {10.1088/0067-0049/219/1/12}, \href
  {https://ui.adsabs.harvard.edu/abs/2015ApJS..219...12A} {219, 12}

\bibitem[\protect\citeauthoryear{{Asplund}, {Grevesse}, {Sauval}  \&
  {Scott}}{{Asplund} et~al.}{2009}]{2009ARA&A..47..481A}
{Asplund} M.,  {Grevesse} N.,  {Sauval} A.~J.,   {Scott} P.,  2009, \mn@doi
  [\araa] {10.1146/annurev.astro.46.060407.145222}, \href
  {https://ui.adsabs.harvard.edu/abs/2009ARA&A..47..481A} {47, 481}

\bibitem[\protect\citeauthoryear{{Bahcall} \& {Salpeter}}{{Bahcall} \&
  {Salpeter}}{1965}]{1965ApJ...142.1677B}
{Bahcall} J.~N.,  {Salpeter} E.~E.,  1965, \mn@doi [\apj] {10.1086/148460},
  \href {https://ui.adsabs.harvard.edu/abs/1965ApJ...142.1677B} {142, 1677}

\bibitem[\protect\citeauthoryear{{Burbidge}, {Lynds}  \& {Burbidge}}{{Burbidge}
  et~al.}{1966}]{1966ApJ...144..447B}
{Burbidge} E.~M.,  {Lynds} C.~R.,   {Burbidge} G.~R.,  1966, \mn@doi [\apj]
  {10.1086/148629}, \href
  {https://ui.adsabs.harvard.edu/abs/1966ApJ...144..447B} {144, 447}

\bibitem[\protect\citeauthoryear{Coles \& Lucchin}{Coles \&
  Lucchin}{2003}]{Coles2003}
Coles P.,  Lucchin P.,  2003, Cosmology: The Origin and Evolution of Cosmic
  Structure.
Wiley, \url {https://books.google.com/books?id=BGYcivB1EtMC}

\bibitem[\protect\citeauthoryear{{Draine}}{{Draine}}{2003}]{2003ARA&A..41..241D}
{Draine} B.~T.,  2003, \mn@doi [\araa]
  {10.1146/annurev.astro.41.011802.094840}, \href
  {https://ui.adsabs.harvard.edu/abs/2003ARA&A..41..241D} {41, 241}

\bibitem[\protect\citeauthoryear{{Draine}}{{Draine}}{2011}]{2011piim.book.....D}
{Draine} B.~T.,  2011, {Physics of the Interstellar and Intergalactic Medium}

\bibitem[\protect\citeauthoryear{{El{\'i}asd{\'o}ttir}
  et~al.,}{{El{\'i}asd{\'o}ttir} et~al.}{2009}]{2009ApJ...697.1725E}
{El{\'i}asd{\'o}ttir} {\'A}.,  et~al., 2009, \mn@doi [\apj]
  {10.1088/0004-637X/697/2/1725}, \href
  {https://ui.adsabs.harvard.edu/abs/2009ApJ...697.1725E} {697, 1725}

\bibitem[\protect\citeauthoryear{{Ellison} et~al.,}{{Ellison}
  et~al.}{2006}]{2006MNRAS.372L..38E}
{Ellison} S.~L.,  et~al., 2006, \mn@doi [\mnras]
  {10.1111/j.1745-3933.2006.00221.x}, \href
  {https://ui.adsabs.harvard.edu/abs/2006MNRAS.372L..38E} {372, L38}

\bibitem[\protect\citeauthoryear{{Fitzpatrick} \& {Massa}}{{Fitzpatrick} \&
  {Massa}}{1990}]{1990ApJS...72..163F}
{Fitzpatrick} E.~L.,  {Massa} D.,  1990, \mn@doi [\apjs] {10.1086/191413},
  \href {https://ui.adsabs.harvard.edu/abs/1990ApJS...72..163F} {72, 163}

\bibitem[\protect\citeauthoryear{{Fitzpatrick} \& {Massa}}{{Fitzpatrick} \&
  {Massa}}{2007}]{2007ApJ...663..320F}
{Fitzpatrick} E.~L.,  {Massa} D.,  2007, \mn@doi [\apj] {10.1086/518158}, \href
  {https://ui.adsabs.harvard.edu/abs/2007ApJ...663..320F} {663, 320}

\bibitem[\protect\citeauthoryear{{Foreman-Mackey}, {Hogg}, {Lang}  \&
  {Goodman}}{{Foreman-Mackey} et~al.}{2013}]{2013PASP..125..306F}
{Foreman-Mackey} D.,  {Hogg} D.~W.,  {Lang} D.,   {Goodman} J.,  2013, \mn@doi
  [\pasp] {10.1086/670067}, \href
  {https://ui.adsabs.harvard.edu/abs/2013PASP..125..306F} {125, 306}

\bibitem[\protect\citeauthoryear{{Gordon}, {Clayton}, {Misselt}, {Landolt}  \&
  {Wolff}}{{Gordon} et~al.}{2003}]{2003ApJ...594..279G}
{Gordon} K.~D.,  {Clayton} G.~C.,  {Misselt} K.~A.,  {Landolt} A.~U.,   {Wolff}
  M.~J.,  2003, \mn@doi [\apj] {10.1086/376774}, \href
  {https://ui.adsabs.harvard.edu/abs/2003ApJ...594..279G} {594, 279}

\bibitem[\protect\citeauthoryear{{Guber} \& {Richter}}{{Guber} \&
  {Richter}}{2016}]{2016A&A...591A.137G}
{Guber} C.~R.,  {Richter} P.,  2016, \mn@doi [\aap]
  {10.1051/0004-6361/201628466}, \href
  {https://ui.adsabs.harvard.edu/abs/2016A&A...591A.137G} {591, A137}

\bibitem[\protect\citeauthoryear{{Guber}, {Richter}  \& {Wendt}}{{Guber}
  et~al.}{2018}]{2018A&A...609A..85G}
{Guber} C.~R.,  {Richter} P.,   {Wendt} M.,  2018, \mn@doi [\aap]
  {10.1051/0004-6361/201730984}, \href
  {https://ui.adsabs.harvard.edu/abs/2018A&A...609A..85G} {609, A85}

\bibitem[\protect\citeauthoryear{{Heintz} et~al.,}{{Heintz}
  et~al.}{2019}]{2019MNRAS.486.2063H}
{Heintz} K.~E.,  et~al., 2019, \mn@doi [\mnras] {10.1093/mnras/stz1012}, \href
  {https://ui.adsabs.harvard.edu/abs/2019MNRAS.486.2063H} {486, 2063}

\bibitem[\protect\citeauthoryear{{Jiang}, {Ge}, {Zhou}, {Wang}  \&
  {Wang}}{{Jiang} et~al.}{2011}]{2011ApJ...732..110J}
{Jiang} P.,  {Ge} J.,  {Zhou} H.,  {Wang} J.,   {Wang} T.,  2011, \mn@doi
  [\apj] {10.1088/0004-637X/732/2/110}, \href
  {https://ui.adsabs.harvard.edu/abs/2011ApJ...732..110J} {732, 110}

\bibitem[\protect\citeauthoryear{{Jiang} et~al.,}{{Jiang}
  et~al.}{2013}]{2013AJ....145..157J}
{Jiang} P.,  et~al., 2013, \mn@doi [\aj] {10.1088/0004-6256/145/6/157}, \href
  {https://ui.adsabs.harvard.edu/abs/2013AJ....145..157J} {145, 157}

\bibitem[\protect\citeauthoryear{{Junkkarinen}, {Cohen}, {Beaver}, {Burbidge},
  {Lyons}  \& {Madejski}}{{Junkkarinen} et~al.}{2004}]{2004ApJ...614..658J}
{Junkkarinen} V.~T.,  {Cohen} R.~D.,  {Beaver} E.~A.,  {Burbidge} E.~M.,
  {Lyons} R.~W.,   {Madejski} G.,  2004, \mn@doi [\apj] {10.1086/423777}, \href
  {https://ui.adsabs.harvard.edu/abs/2004ApJ...614..658J} {614, 658}

\bibitem[\protect\citeauthoryear{{Ledoux}, {Noterdaeme}, {Petitjean}  \&
  {Srianand}}{{Ledoux} et~al.}{2015}]{2015A&A...580A...8L}
{Ledoux} C.,  {Noterdaeme} P.,  {Petitjean} P.,   {Srianand} R.,  2015, \mn@doi
  [\aap] {10.1051/0004-6361/201424122}, \href
  {https://ui.adsabs.harvard.edu/abs/2015A&A...580A...8L} {580, A8}

\bibitem[\protect\citeauthoryear{{Ma} et~al.,}{{Ma}
  et~al.}{2015}]{2015MNRAS.454.1751M}
{Ma} J.,  et~al., 2015, \mn@doi [\mnras] {10.1093/mnras/stv2073}, \href
  {https://ui.adsabs.harvard.edu/abs/2015MNRAS.454.1751M} {454, 1751}

\bibitem[\protect\citeauthoryear{{Ma}, {Ge}, {Zhao}, {Prochaska}, {Zhang}, {Ji}
   \& {Schneider}}{{Ma} et~al.}{2017}]{2017MNRAS.472.2196M}
{Ma} J.,  {Ge} J.,  {Zhao} Y.,  {Prochaska} J.~X.,  {Zhang} S.,  {Ji} T.,
  {Schneider} D.~P.,  2017, \mn@doi [\mnras] {10.1093/mnras/stx2117}, \href
  {https://ui.adsabs.harvard.edu/abs/2017MNRAS.472.2196M} {472, 2196}

\bibitem[\protect\citeauthoryear{{Ma} et~al.,}{{Ma}
  et~al.}{2018}]{2018MNRAS.474.4870M}
{Ma} J.,  et~al., 2018, \mn@doi [\mnras] {10.1093/mnras/stx3123}, \href
  {https://ui.adsabs.harvard.edu/abs/2018MNRAS.474.4870M} {474, 4870}

\bibitem[\protect\citeauthoryear{{Morton}}{{Morton}}{1991}]{1991ApJS...77..119M}
{Morton} D.~C.,  1991, \mn@doi [\apjs] {10.1086/191601}, \href
  {https://ui.adsabs.harvard.edu/abs/1991ApJS...77..119M} {77, 119}

\bibitem[\protect\citeauthoryear{{Motta} et~al.,}{{Motta}
  et~al.}{2002}]{2002ApJ...574..719M}
{Motta} V.,  et~al., 2002, \mn@doi [\apj] {10.1086/341118}, \href
  {https://ui.adsabs.harvard.edu/abs/2002ApJ...574..719M} {574, 719}

\bibitem[\protect\citeauthoryear{{Nestor}, {Pettini}, {Hewett}, {Rao}  \&
  {Wild}}{{Nestor} et~al.}{2008}]{2008MNRAS.390.1670N}
{Nestor} D.~B.,  {Pettini} M.,  {Hewett} P.~C.,  {Rao} S.,   {Wild} V.,  2008,
  \mn@doi [\mnras] {10.1111/j.1365-2966.2008.13857.x}, \href
  {https://ui.adsabs.harvard.edu/abs/2008MNRAS.390.1670N} {390, 1670}

\bibitem[\protect\citeauthoryear{{Pettini}, {Boksenberg}  \&
  {Hunstead}}{{Pettini} et~al.}{1990}]{1990ApJ...348...48P}
{Pettini} M.,  {Boksenberg} A.,   {Hunstead} R.~W.,  1990, \mn@doi [\apj]
  {10.1086/168212}, \href
  {https://ui.adsabs.harvard.edu/abs/1990ApJ...348...48P} {348, 48}

\bibitem[\protect\citeauthoryear{{Quiret} et~al.,}{{Quiret}
  et~al.}{2016}]{2016MNRAS.458.4074Q}
{Quiret} S.,  et~al., 2016, \mn@doi [\mnras] {10.1093/mnras/stw524}, \href
  {https://ui.adsabs.harvard.edu/abs/2016MNRAS.458.4074Q} {458, 4074}

\bibitem[\protect\citeauthoryear{{Richter}, {Krause}, {Fechner}, {Charlton}  \&
  {Murphy}}{{Richter} et~al.}{2011}]{2011A&A...528A..12R}
{Richter} P.,  {Krause} F.,  {Fechner} C.,  {Charlton} J.~C.,   {Murphy} M.~T.,
   2011, \mn@doi [\aap] {10.1051/0004-6361/201015566}, \href
  {https://ui.adsabs.harvard.edu/abs/2011A&A...528A..12R} {528, A12}

\bibitem[\protect\citeauthoryear{{Sardane}, {Turnshek}  \& {Rao}}{{Sardane}
  et~al.}{2014}]{2014MNRAS.444.1747S}
{Sardane} G.~M.,  {Turnshek} D.~A.,   {Rao} S.~M.,  2014, \mn@doi [\mnras]
  {10.1093/mnras/stu1554}, \href
  {https://ui.adsabs.harvard.edu/abs/2014MNRAS.444.1747S} {444, 1747}

\bibitem[\protect\citeauthoryear{{Sardane}, {Turnshek}  \& {Rao}}{{Sardane}
  et~al.}{2015}]{2015MNRAS.452.3192S}
{Sardane} G.~M.,  {Turnshek} D.~A.,   {Rao} S.~M.,  2015, \mn@doi [\mnras]
  {10.1093/mnras/stv1506}, \href
  {https://ui.adsabs.harvard.edu/abs/2015MNRAS.452.3192S} {452, 3192}

\bibitem[\protect\citeauthoryear{{Savage} \& {Sembach}}{{Savage} \&
  {Sembach}}{1996}]{1996ARA&A..34..279S}
{Savage} B.~D.,  {Sembach} K.~R.,  1996, \mn@doi [\araa]
  {10.1146/annurev.astro.34.1.279}, \href
  {https://ui.adsabs.harvard.edu/abs/1996ARA&A..34..279S} {34, 279}

\bibitem[\protect\citeauthoryear{{Schlegel}, {Finkbeiner}  \&
  {Davis}}{{Schlegel} et~al.}{1998}]{1998ApJ...500..525S}
{Schlegel} D.~J.,  {Finkbeiner} D.~P.,   {Davis} M.,  1998, \mn@doi [\apj]
  {10.1086/305772}, \href
  {https://ui.adsabs.harvard.edu/abs/1998ApJ...500..525S} {500, 525}

\bibitem[\protect\citeauthoryear{{Wang}, {Hall}, {Ge}, {Li}  \&
  {Schneider}}{{Wang} et~al.}{2004}]{2004ApJ...609..589W}
{Wang} J.,  {Hall} P.~B.,  {Ge} J.,  {Li} A.,   {Schneider} D.~P.,  2004,
  \mn@doi [\apj] {10.1086/421240}, \href
  {https://ui.adsabs.harvard.edu/abs/2004ApJ...609..589W} {609, 589}

\bibitem[\protect\citeauthoryear{Waskom}{Waskom}{2021}]{Waskom2021}
Waskom M.~L.,  2021, \mn@doi [Journal of Open Source Software]
  {10.21105/joss.03021}, 6, 3021

\bibitem[\protect\citeauthoryear{{Welty}, {Hobbs}, {Lauroesch}, {Morton},
  {Spitzer}  \& {York}}{{Welty} et~al.}{1999}]{1999ApJS..124..465W}
{Welty} D.~E.,  {Hobbs} L.~M.,  {Lauroesch} J.~T.,  {Morton} D.~C.,  {Spitzer}
  L.,   {York} D.~G.,  1999, \mn@doi [\apjs] {10.1086/313263}, \href
  {https://ui.adsabs.harvard.edu/abs/1999ApJS..124..465W} {124, 465}

\bibitem[\protect\citeauthoryear{{Wild} \& {Hewett}}{{Wild} \&
  {Hewett}}{2005}]{2005MNRAS.361L..30W}
{Wild} V.,  {Hewett} P.~C.,  2005, \mn@doi [\mnras]
  {10.1111/j.1745-3933.2005.00058.x}, \href
  {https://ui.adsabs.harvard.edu/abs/2005MNRAS.361L..30W} {361, L30}

\bibitem[\protect\citeauthoryear{{Wild}, {Hewett}  \& {Pettini}}{{Wild}
  et~al.}{2006}]{2006MNRAS.367..211W}
{Wild} V.,  {Hewett} P.~C.,   {Pettini} M.,  2006, \mn@doi [\mnras]
  {10.1111/j.1365-2966.2005.09935.x}, \href
  {https://ui.adsabs.harvard.edu/abs/2006MNRAS.367..211W} {367, 211}

\bibitem[\protect\citeauthoryear{{Wolfe}, {Turnshek}, {Smith}  \&
  {Cohen}}{{Wolfe} et~al.}{1986}]{1986ApJS...61..249W}
{Wolfe} A.~M.,  {Turnshek} D.~A.,  {Smith} H.~E.,   {Cohen} R.~D.,  1986,
  \mn@doi [\apjs] {10.1086/191114}, \href
  {https://ui.adsabs.harvard.edu/abs/1986ApJS...61..249W} {61, 249}

\bibitem[\protect\citeauthoryear{{Xia}, {Ge}, {Willis}  \& {Zhao}}{{Xia}
  et~al.}{2022}]{2022MNRAS}
{Xia} I.,  {Ge} J.,  {Willis} K.,   {Zhao} Y.,  2022, \mnras, submitted

\bibitem[\protect\citeauthoryear{{York} et~al.,}{{York}
  et~al.}{2006}]{2006MNRAS.367..945Y}
{York} D.~G.,  et~al., 2006, \mn@doi [\mnras]
  {10.1111/j.1365-2966.2005.10018.x}, \href
  {https://ui.adsabs.harvard.edu/abs/2006MNRAS.367..945Y} {367, 945}

\bibitem[\protect\citeauthoryear{{Zafar}, {Watson}, {Fynbo}, {Malesani},
  {Jakobsson}  \& {de Ugarte Postigo}}{{Zafar}
  et~al.}{2011}]{2011A&A...532A.143Z}
{Zafar} T.,  {Watson} D.,  {Fynbo} J.~P.~U.,  {Malesani} D.,  {Jakobsson} P.,
  {de Ugarte Postigo} A.,  2011, \mn@doi [\aap] {10.1051/0004-6361/201116663},
  \href {https://ui.adsabs.harvard.edu/abs/2011A&A...532A.143Z} {532, A143}

\bibitem[\protect\citeauthoryear{{Zhao}, {Ge}, {Yuan}, {Zhao}, {Wang}  \&
  {Li}}{{Zhao} et~al.}{2019}]{2019MNRAS.487..801Z}
{Zhao} Y.,  {Ge} J.,  {Yuan} X.,  {Zhao} T.,  {Wang} C.,   {Li} X.,  2019,
  \mn@doi [\mnras] {10.1093/mnras/stz1197}, \href
  {https://ui.adsabs.harvard.edu/abs/2019MNRAS.487..801Z} {487, 801}

\bibitem[\protect\citeauthoryear{{Zhu} \& {M{\'e}nard}}{{Zhu} \&
  {M{\'e}nard}}{2013}]{2013ApJ...770..130Z}
{Zhu} G.,  {M{\'e}nard} B.,  2013, \mn@doi [\apj]
  {10.1088/0004-637X/770/2/130}, \href
  {https://ui.adsabs.harvard.edu/abs/2013ApJ...770..130Z} {770, 130}

\bibitem[\protect\citeauthoryear{{Zych}, {Murphy}, {Pettini}, {Hewett},
  {Ryan-Weber}  \& {Ellison}}{{Zych} et~al.}{2007}]{2007MNRAS.379.1409Z}
{Zych} B.~J.,  {Murphy} M.~T.,  {Pettini} M.,  {Hewett} P.~C.,  {Ryan-Weber}
  E.~V.,   {Ellison} S.~L.,  2007, \mn@doi [\mnras]
  {10.1111/j.1365-2966.2007.12015.x}, \href
  {https://ui.adsabs.harvard.edu/abs/2007MNRAS.379.1409Z} {379, 1409}

\makeatother
\end{thebibliography}

\bsp	% typesetting comment
\label{lastpage}
\end{document}